\documentclass[usenatbib]{mnras}
\usepackage{newtxtext,newtxmath}

\usepackage[T1]{fontenc}

\DeclareRobustCommand{\VAN}[3]{#2}
\let\VANthebibliography\thebibliography
\def\thebibliography{\DeclareRobustCommand{\VAN}[3]{##3}\VANthebibliography}

\usepackage{graphicx}  
\usepackage{amsmath}  
\usepackage{float}
\usepackage[dvipsnames]{xcolor}
\usepackage{hyperref}
\usepackage{orcidlink}
\usepackage[most]{tcolorbox}
\usepackage{subcaption}
\usepackage{enumitem}
\usepackage{soul}
\numberwithin{equation}{section}

\newtcolorbox{caveat}{colback=red!5!bg,colframe=orange!75!black,boxrule=0.45mm,coltext=fg}
\newtcolorbox{note}{colback=blue!10!white,colframe=blue!75!black,boxrule=0.45mm,coltext=black, sharp corners=downhill}
\newtcolorbox{highlight}{colback=yellow!10!white,colframe=yellow!75!black,boxrule=0.45mm,coltext=black, sharp corners=downhill,top=0mm,bottom=0mm,left=1mm,right=1mm}
\newtcolorbox{warning}[1][]{
  colback=yellow!5!white,
  colframe=yellow!75!black,
  top=4mm,
  fonttitle=\bfseries,
  colbacktitle=yellow!75!black,enhanced,
  attach boxed title to top left={yshift=-4mm, xshift=2mm},
  title={\faExclamationTriangle\quad}, 
  #1
}
\newtcolorbox{swarning}{ 
  colback=yellow!20!white,
  colframe=yellow!80!black,
  coltext=black,
  sharp corners=west,
  fonttitle=\bfseries,
  boxrule=0.5mm,
  leftrule=6mm,
  toprule=0pt,
  bottomrule=0pt,
  enhanced,
  overlay={
    \node[anchor=center, text=white, inner sep=1pt, circle] at ([xshift=3mm]frame.west) {\bfseries\LARGE !};
  }
}

\soulregister{\citet}{7}
\soulregister{\citep}{7}
\soulregister{\cite}{7}
\soulregister{\citealt}{7}
\soulregister{\ref}{7}
\setstcolor{red} 
\sethlcolor{SpringGreen} 

\newcommand{\change}[2]{{\st{#1}} {\hl{#2}}}

\newcommand{\MSUN}{\rm M_{\odot}}

\newcommand{\MTWOC}{M_{\rm 200c}}
\newcommand{\RTWOC}{R_{\rm 200c}}

\newcommand{\MFC}{M_{\rm 500c}}
\newcommand{\RFC}{R_{\rm 500c}}

\newcommand{\svt}{\sigma_{v,\rm{Turb}}}

\newcommand{\svtot}{\sigma_{v,\rm{Total}}}

\title[Bulk vs. turbulent ICM motions with TNG-Cluster]{Bulk vs. turbulent motions at the centres of galaxy clusters:\\ AGN-driven turbulence according to TNG-Cluster}

\author[B. Saha et al.]{Bipradeep Saha \orcidlink{0000-0002-2329-7340},$^{1,2}$\thanks{E-mail: bisaha@mpia.de}
  Annalisa Pillepich\orcidlink{0000-0003-1065-927},$^{1}$
  Joey Braspenning\orcidlink{0009-0003-3956-4890},$^1$
  Marine Prunier\orcidlink{0009-0003-0932-2487},$^{1, 3}$
  \newauthor
  Dimitris Chatzigiannakis\orcidlink{0009-0008-3247-9489},$^{1,2}$
  and Dylan Nelson\orcidlink{0000-0001-8421-5890}$^{4,5}$
  \\
  $^1$Max-Planck-Institut f{\"u}r Astronomie, K{\"o}nigstuhl 17, D-69117 Heidelberg, Germany\\
  $^2$Fakult\"at f\"ur Physik und Astronomie, Universit\"at Heidelberg, Im Neuenheimer Feld 226, 69120 Heidelberg, Germany\\
  $^3$Département de Physique, Université de Montréal, Succ. Centre-Ville, Montréal, Québec, H3C 3J7, Canada\\
  $^4$Universit\"at Heidelberg, Zentrum f\"ur Astronomie, ITA, Albert-Ueberle-Str. 2, D-69120 Heidelberg, Germany\\
  $^5$Universit\"at Heidelberg, Interdisziplin\"ares Zentrum f\"ur Wissenschaftliches Rechnen, INF 205, 69120 Heidelberg, Germany
}

\date{Accepted XXX. Received YYY; in original form ZZZ}

\pubyear{\the\year{}}

\begin{document}
\label{firstpage}
\pagerange{\pageref{firstpage}--\pageref{lastpage}}
\maketitle

\begin{abstract}
  {
    The highly dynamic intracluster medium (ICM) influences cluster thermodynamic evolution and probes key physical processes. Quantifying the non-thermal motions is therefore essential for understanding cluster physics and interpreting high spectral-resolution X-ray observations from telescopes like {\it XRISM}.
    We quantify bulk and turbulent gas motions in 352 galaxy clusters at $z=0$ (${\rm M_{200c}=10^{14.3-15.4}\, M_\odot}$) from the TNG-Cluster suite of magneto-hydrodynamical galaxy simulations. We use a multi-scale filtering Reynolds decomposition to separate total gas velocities into bulk (coherent) and turbulent (small-scale fluctuations) components.  We primarily focus on the hot X-ray emitting gas in the central core regions.
    According to TNG-Cluster, majority of the ICM has subsonic turbulence but with broad velocity distributions reaching $\mathcal{M}_{\rm Turb}\sim 10$ and large cluster-to-cluster variations. In cluster centres, turbulence contributes less than half of the total velocity dispersion $(\sigma_{v\rm,Turb } \sim 0.5 ~\sigma_{v,\rm Total})$ for most clusters, with typical turbulent velocity dispersions of $50-75$ km s$^{-1}$ across the mass range, and with sub per cent levels of turbulent pressure support. Clusters that are strong cool cores, or have X-ray cavities, or experienced recent SMBH feedback energy injections exhibit systematically larger turbulent velocity dispersions and more prominent turbulent velocity tails. On average, the turbulent velocity dispersion peaks in cluster centres, decreases slightly to a minimum at $0.1-0.2 \, R_{\rm500c}$, then rises again.
    Our analysis shows that SMBH feedback is a key driver of turbulence in cluster cores, generating strong but short-lived motion alongside high-velocity outflows. It also calls for caution for interpreting {\it XRISM} observations.
  }
\end{abstract}




\section{Introduction}
\label{sec:intro}
With the advent of new high spectral resolution X-ray telescopes, such as {\it XRISM} \citep{2025_XRISM,2020_XRISM_whitepaper} and the future {\it NewAthena} \citep{2024_NewAthena_whitepaper}, we can study the hot \change{gas /}{} plasma (temperature $ \sim 10^{7-8} $ K) in the {Universe} with very high precision. The majority of this hot plasma resides in the intracluster medium (ICM) of galaxy clusters -- hereafter clusters. Clusters are virialized objects in the Universe, with total halo masses ranging from $ 10^{14-15.5} \MSUN $ and in the standard cosmological paradigm they grow hierarchically via mergers and accretion of galaxies \citep[see e.g.,][]{DeLucia_2007,Naab_2009,vanDokkum_2010}. The ICM fills the space between galaxies in a cluster and makes up about $10-15$ per cent of the total mass of the cluster {and about $\sim 85-90$ per cent of the cluster's baryonic mass}. It is observable {by} its X-ray emission \citep[see e.g.,][]{1975_Gull,1999_Cavaliere} due to thermal bremsstrahlung radiation  {with additional contributions from line emission due to transitions in highly ionized heavy elements (e.g., Fe)}.

Crucially, observational data reveals that the hot ICM is not static. Gas kinematics along the line-of-sight can be directly probed by measuring the Doppler shift and broadening of emission lines  to obtain, for example, the velocity dispersion $ (\sigma _v) $  of the gas \citep[e.g.,][]{Inogamov_2003,Churazov_2004,Zhuravleva_2012,ZuHone_2018}. Recent observational campaigns using high-resolution X-ray spectroscopy have measured the turbulent velocity dispersion in the cores of several  galaxy clusters, revealing low levels of turbulence. For instance, observations of the Perseus cluster by the {\it Hitomi} satellite and subsequently {\it XRISM} indicate a velocity dispersion of roughly $ 160-200$ km s$^{-1}$, contributing to less than $5$ per cent of the {total} pressure support \citep{Hitomi_2016, XRISM_Perseus_2025}. Similar low levels of turbulence have been inferred in other systems, such as Abell 2029, Coma, Abell 2319, and Ophiuchus \citep{XRISM_A2319_2025,XRISM_A2029_2025,XRISM_Coma_2025,XRISM_Ophiuchus_2025}. Alternative observational techniques, such as measuring {X-ray} surface brightness fluctuations \citep[e.g.,][]{Churazov_2012,Zhuravleva_2014,Gaspari_2014}, utilizing the thermal Sunyaev-Zel'dovich (tSZ) effect \citep{Khatri_2016}, and analysing deviations from hydrostatic equilibrium \citep{Eckart_2019}, generally corroborate that non-thermal pressure support constitutes only a modest fraction of the total energy budget in the centres of galaxy clusters.

These observational findings suggest {a potential} tension with theoretical expectations {and simulation results, which appear to predict, at least qualitatively and for the average cluster, larger contributions from non-thermal motions than those recently inferred observationally, as discussed below}.

In the standard cosmological paradigm, it is expected that the ICM is constantly stirred by diverse astrophysical processes. Theoretical models suggest that {structure formation -- including cosmic accretion from the cosmic web and galaxy mergers} -- drive significant internal motions, heating the gas via shocks and adiabatic compression \citep[e.g.,][]{Vazza_2009b,Biffi_2016,Simionescu_2019}. Furthermore, it is broadly accepted that energetic feedback from supermassive black holes (SMBHs) at the centres of brightest cluster galaxies (BCGs) is required to prevent runaway cooling flows \citep[e.g.,][]{Fabian_1994, Ciotti_2001, Peterson_2006, McNamara_2007}. This active galactic nucleus (AGN) feedback is theoretically expected to inject substantial kinetic energy into the surrounding gas, efficiently stirring the ICM \citep[e.g.,][]{Fabian_2012, Gaspari_2020}.

To understand the nature of gas motions in galaxy clusters and to interpret observational kinematic data, first-principle numerical simulations of ICM turbulence are essential, especially those that account for the cosmological hierarchical growth of structure and cosmological mass accretion. Cosmological simulations of galaxy clusters including gravity and (magneto-)hydrodynamics (MHD) have shown that gas motions are ubiquitous and can significantly contribute to the energy and pressure budget of the ICM, even in the absence of gas cooling, galactic astrophysics and galaxy feedback processes \citep[e.g.,][]{Norman_1999,Dolag_2005,Lau_2009,Vazza_2009b,Vazza_2011}. These have often predicted typical non-thermal pressure support of $5-30$ per cent of the thermal pressure, depending on cluster mass and radius \citep{Lau_2009,Nelson_2014, Vazza_2018, Angelinelli_2020, Sayers_2021, Groth_2025a}. Neglecting this non-thermal support has been shown to lead to biases in hydrostatic mass estimates \citep[e.g.,][]{Nagai_2007,Nagai_2013,Shi_2014,Shi_2015}, with far reaching consequences beyond ICM astrophysics.

{However, accurately simulating non-thermal motions in self-consistent models of clusters and their ICM is non trivial. Non-thermal motions include, a priori, bulk (i.e. coherent) flows such as cosmological gas inflows, sloshing patterns, SMBH-driven outflows and shear motions, in addition to random eddy-like velocity fluctuations, as well as shocks and sound waves. Accurately simulating the latter, i.e. turbulence, is a notoriously difficult computational problem. Capturing the full-physical picture requires resolving the turbulent energy cascade from the large-scale injection mechanisms down through the inertial range, which poses severe resolution constraints for large-volume cosmological simulations (e.g. \citealt{howes_inertial_2008,howes_weakened_2011,Bauer_2012,federrath_2026}). Consequently} predictions of turbulence and non-thermal motions in clusters may depend on the underlying hydrodynamical schemes \citep[e.g.,][]{Dolag_2004, Groth_2025a}. Furthermore, they are highly sensitive to the physical processes included in the models, for example whether active galactic nucleus (AGN) feedback is included \citep{Sotira_2025}. Additionally, previous numerical investigations have often been limited by the lack of large, statistically-representative samples of simulated galaxy clusters that simultaneously possess the resolution and model complexity needed to resolve core dynamics and the physics of cluster galaxies within a cosmological context. These processes are naturally captured in the uniform-volume cosmological simulations of realistic galaxy populations developed over the last decade, such as Illustris, EAGLE, and IllustrisTNG, with baryonic mass resolution of $10^{6-7}\MSUN$. However, these simulations do not cover sufficiently large volumes to yield more than a few truly massive ($10^{15}\MSUN$) galaxy clusters.

{Even when the turbulent cascade is adequately resolved in simulations, extracting physically-meaningful conclusions presents a conceptual challenge: identifying and isolating turbulent motions from the complex total velocity field. Observationally, X-ray line broadening captures all line-of-sight velocity variations indiscriminately. However, these non-thermal motions encompass a complex superposition of velocity components across varying spatial scales and, crucially, bulk flows and isotropic turbulence have vastly different implications for energy dissipation rates, the coupling efficiency of AGN feedback, and biases in hydrostatic mass estimates. Therefore, to meaningfully interpret observational line widths and isolate the true chaotic cascades, we must robustly disentangle small-scale turbulence from large-scale bulk flows.}
Early numerical approaches based on fixed spatial filtering scales revealed significant sensitivity to the choice of filter scale \citep{Vazza_2012}, motivating the development of adaptive multi-scale filtering techniques \citep{Vazza_2017}.

\begin{figure*}
  \includegraphics[width=0.9\textwidth]{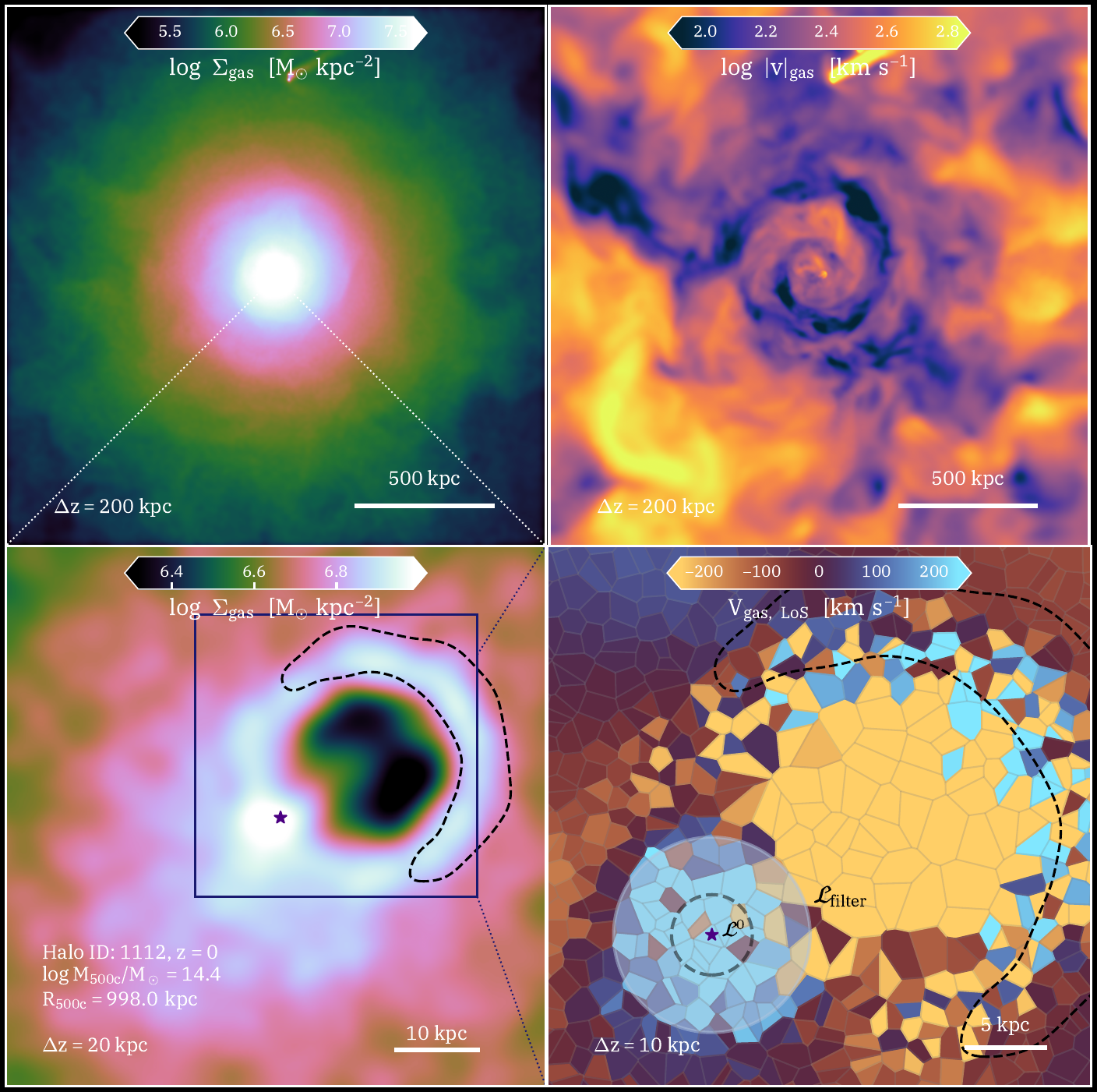}
  \caption{{\bf Functioning of the multi-scale filtering Reynolds Decomposition (RD) used throughout this work to separate bulk vs. turbulent gas motions and applied to the ICM of cosmological MHD simulations of galaxy clusters from the TNG-Cluster suite.} Here we showcase the application of our method to one example simulated cluster, projected along the $z$ axis of the simulation box. In the {\bf Top}, we show projected gas density (left) and total velocity (right) in a 200 kpc-thick slice centred on the cluster, with a field of view of $2\,\RFC$ per side. In the {\bf Bottom}, we progressively zoom in to smaller scales and show projected gas density in a 20 kpc-thick slice with a 67 kpc field of view (left) and the line-of-sight velocities for individual Voronoi gas cells in a 10 kpc-thick slice with a 40 kpc field of view (right).
    In the bottom panels, black contours mark regions of high subsonic flow (local Mach numbers in the range 0.25 – 0.9). We use these contours as a visual guide to highlight the kinematics surrounding a rising X-ray cavity and its associated shock front (the cavity and shock were formally identified in this system by \citet{Prunier_2025a})
  The TNG-Cluster suite naturally returns complex ICM kinematics, with  {many contributing processes} across varying spatial scales in any single cluster. The multi-scale RD identifies, for each gas cell, a filtering scale for turbulence -- i.e. the spatial scale beyond which the turbulent velocities do not change beyond a certain relative difference -- we call this turbulence filtering scale and denote as $\mathcal{L}_{\rm filter}$. Illustrated for the central gas cell of this cluster (purple star) are the initial searching filtering scale (dashed, $\mathcal{L}^{0}$) and the final converged filtering scale (solid, $\mathcal{L}_{\rm filter}$). The algorithm hence provides adaptive filtering scales for turbulence, one for each gas cell, and these can be larger than the adopted field of view.}
  \label{fig:RD_Schematic}
\end{figure*}

In this paper, we build on the idea of a Reynolds decomposition based on multi-scale filtering to distinguish and characterize bulk and turbulent motions in the ICM of the TNG-Cluster simulation suite \citep{Nelson_2024}. Comprising 352 cosmological zoom-in MHD simulations, TNG-Cluster represents the largest sample of high-resolution ($10^7\MSUN$ baryonic mass) simulated galaxy clusters to date that includes galaxy-formation physics and spans a wide mass range of $10^{14.3-15.4}\MSUN$ at $z=0$. TNG-Cluster complements even larger suites of simulated galaxy clusters that, however, operate at one to two orders of magnitude lower numerical resolution, such as MUSIC \citep{Sembolini_2013}, the Three Hundred Project \citep{Cui_2018,Cui_2022}, Magneticum-1mr \citep{Dolag_2016}, and more recently FLAMINGO \citep{Schaye_2023}.

Crucially, in addition to hosting a large and diverse population of galaxy clusters, TNG-Cluster is based on the IllustrisTNG galaxy-formation model \citep{Weinberger_2017, Pillepich_2018a}, which has been extensively tested and studied both in terms of galaxy properties and populations and in the context of galaxy clusters. TNG-Cluster reproduces a broad diversity of observed phenomena, including strong, weak, and non-cool cores \citep{Lehle_2024, Lehle_2025}, a non-negligible cool gas component \citep{Rohr_2024, Staffehl_2025}, diverse radio relic populations \citep{Lee_2024, Lee_2026}, and rich X-ray morphologies that are reminiscent of, and even quantitatively consistent with, observations, including shocks \citep{Prunier_2025c} and X-ray cavities \citep{Prunier_2025a, Prunier_2025b}, driven by the kinetic feedback from low-accreting supermassive black holes (SMBHs).

Particularly relevant for this analysis, the levels of gas velocity dispersion predicted by TNG-Cluster are broadly in the ballpark of observed values, especially for Perseus-like clusters \citep{Truong_2024}, albeit with some potential biases across the cluster population (\citealt{XRISM-Truong_2025} but see \citealt{Chatzigiannakis_2026}). Here, we aim to quantify the intrinsic levels of turbulence in the ICM, disentangling them from large-scale bulk flows, in order to determine the {solely} turbulent velocity dispersion of the ICM and to understand its dependence on cluster properties and underlying physical drivers.

Therefore, in Section~\ref{sec:methods}, we describe the TNG-Cluster simulation suite and our methodology for Reynolds decomposition, including sensitivity tests and operational definitions. In Section~\ref{sec:results}, we present our main results on the ICM kinematics in cluster centres, including the complexity of velocity fields, velocity distributions, the relationship between total and turbulent velocity dispersions, and the physical drivers of turbulence. In Section~\ref{sec:disc}, we discuss the implications of our findings, compare with existing results, address limitations of our approach, and provide interpretation guidance for observational results. Finally, in Section~\ref{sec:conclusions}, we summarize our conclusions.

\section{Methodology and TNG-Cluster}
\label{sec:methods}

\subsection{The TNG-Cluster simulation suite}
\label{sec:methods_TNG-Cluster}
The TNG-Cluster simulation suite \citep{Nelson_2024} is a set of 352 high-resolution magneto-hydrodynamical (MHD) cosmological zoom-in simulations of galaxy clusters in the standard $\Lambda$CDM model. It is a spin-off of IllustrisTNG \citep[hereafter TNG; see ][]{Pillepich_2018b,Nelson_2018,Springel_2018,Naiman_2018,Marinacci_2018,Pillepich_2019_TNG50, Nelson_2019,Nelson_2019_TNG50}, run with the moving-mesh code \texttt{AREPO} \citep{Springel_2010}.

The TNG-Cluster suite extends TNG300 (TNG300-1) of IllustrisTNG by improving the statistics of high-mass haloes. Target haloes for the zoom-in re-simulations were selected from a large dark matter-only simulation of a periodic box of $\sim 1$ comoving Gpc per side. Selection was based solely on halo mass at $z=0$: all systems with $\log M_{\rm 200c}/\MSUN > 15.0$ were included, while those in the range $14.2 < \log M_{\rm 200c}/\MSUN < 15.0$ were randomly sampled in $0.1$ dex bins. This strategy compensates for the decline of high-mass haloes in the TNG300 volume and yields a more uniform mass distribution \citep[see Fig.~1 of][]{Nelson_2024}.

The TNG-Cluster suite  has the same resolution as TNG300, i.e $ m_{\rm{gas, target}} = 1.2 \times 10^7 \MSUN$ and $ m_{\rm{DM}} = 6.1 \times 10^7 \MSUN  $. It incorporates the TNG galaxy formation model \citep[described in][]{Weinberger_2017,Pillepich_2018a} that is well validated and includes  much of the physics thought to be of most relevance for galaxy evolution. This includes gas processes such as heating from the UV background and radiative cooling. It incorporates star formation, stellar evolution, and chemical enrichment. The model also accounts for stellar feedback via galactic-scale winds, alongside the seeding, merging, and energetic feedback of {SMBHs}, i.e., {AGNs}. TNG-Cluster also adopts the TNG cosmology: $ \Omega _m = 0.3089, \Omega _b = 0.0486, \Omega _\Lambda=0.6911, H_0=67.74~h~{\rm{ km / s / Mpc }}, \sigma _8=0.8159 $ and $ n_s=0.9667 $ \citep{Planck_2016}.

The haloes are then identified using the standard, friends-of-friends (FoF) algorithm  with a linking length of $ b = 0.2 $, and the sub-structures were identified using \texttt{SUBFIND} routine \citep{Springel_2001} and were linked across different snapshots using the \texttt{SubLink} algorithm \citep{Rodriguez-Gomez_2015}.

\subsection{Focus on cluster centres and operational definitions}
\label{sec:methods_defs}

In this paper, we make use of all the target 352 galaxy clusters of the TNG-Cluster suite, by focusing mostly at low redshift, $z=0$.

{We generally consider} the central regions of the clusters, defined as a sphere with radius $r = 33.5~\rm{kpc}$ centred on the potential minimum of the primary subhalo (usually their brightest cluster galaxy, BCG). This physical aperture corresponds to the field of view (FoV) of a single {\it XRISM} pointing for the Perseus cluster at $z \approx 0.1$.

In this study, we primarily focus on the volume-filling hot ICM, to more directly connect to X-ray observations, but also compare results across different thermodynamic gas phases. Our definition of the different phases are as follows:
\begin{itemize}
  \item Hot (i.e X-ray-emitting gas): $ T> 10^{5.5}~{K} $
  \item Warm: $ 10^{4.5} < T \leq 10^{5.5}~{K} $
  \item Cool: $ T \leq 10^{4.5}~{K} $.
\end{itemize}
Since we do not explicitly model gas colder than $ 10^4~{\rm{K}} $ due to temperature floor in the ISM effective equation of state \citep[see][]{Springel_2003, Pillepich_2018a}, we artificially assign a temperature $ 10^{3}~{\rm{K}} $ to the star-forming gas cells.

We quantify the kinematics of the ICM by using throughout mass-weighted velocities and velocity dispersions (hereafter omitting the weighting):
\begin{align}
  v_{i, w} &= \frac{\sum_{j} w_j v_{i,j} }{\sum_{j} w_j}\\
  \sigma ^2_{i, w} &= \frac{\sum_{j} w_j v_{i, j}^2}{\sum_{j} w_j} - v_{i, w}^2
  \label{eq:sigma}
\end{align}
where $ i $  is indexed over $x, y, z$ components of the velocity field, $ j $ is indexed over all gas cells in the FoV, and $ w_j $ is the mass of the gas cell $ j $. When we compute the 3D velocity dispersion, we use the magnitude of the velocity vector, $ v= \sqrt{v_{x}^2 + v_{y}^2 + v_{z}^2}  $.

As anticipated in Section~\ref{sec:intro}, we do not employ any forward modelling of X-ray observables from TNG-Cluster \citep[but see][]{Prunier_2025a, Prunier_2025b, Prunier_2025c, Chatzigiannakis_2026}, as we aim to provide intrinsic physical context rather than direct mock observations or comparisons to real data. In fact, \cite{Truong_2024} have shown, using end-to-end mock {\it XRISM} observations of the cores of Perseus-like clusters from TNG-Cluster, that the velocity dispersion inferred via X-ray spectral fitting captures, on average, the emission-weighted velocity dispersion well and, in certain cases, underestimates the mass-weighted intrinsic velocity dispersion, with typical errors smaller than 30 per cent. This can be used as a gauge to translate our findings into more observationally-connected results.

Finally, it will be useful to compare the non-thermal energy (i.e. kinetic energy) of the ICM gas to the thermal one, to get an idea of the level of thermal pressure support, e.g. via the ratio of non-thermal to thermal pressure: $ P_{\rm{nt}} / P_{\rm{th}} $. This ratio can be estimated as \citep[e.g.,][]{Vazza_2018}:
\begin{equation}
  \frac{P_{\rm{Non-thermal}}}{P_{\rm{Thermal}}} \frac{}{} = \frac{\rho \sigma^2_{\rm{3D}} / 3}{\rho k_B T / (\mu_e m_p)}
  \label{eq:P_ratios}
\end{equation}
where $\sigma_{\rm{3D}}$ is the three-dimensional velocity dispersion (the factor $1/3$ accounts for isotropy), $k_B$ is the Boltzmann constant, $T$ is the mass-weighted gas temperature, $\mu_e$ is the mean molecular weight per free electron, and $m_p$ is the proton mass. We note that the numerator of Eq.~\ref{eq:P_ratios}, the non-thermal pressure of the gas, scales as the kinetic energy density, i.e. $ \propto \rho v^2$, with $v$ being the gas velocity. However, in observational literature, $v$ is often replaced with what can be inferred observationally, i.e. the velocity dispersion \citep[e.g.,][]{XRISM_Perseus_2025,XRISM_A2029_2025}.

\subsection{Turbulent vs. bulk motions via Reynold's Decomposition}
Due to the absence of a univocal definition of turbulence \citep[see e.g.,][]{Adrian_2000}, separating the velocity field into bulk (i.e coherent) and turbulent (i.e chaotic) -- or in other words, performing a Reynold Decomposition (RD) -- is non-trivial. Earlier {approaches to perform RD in cosmological simulations} consisted of filtering out the turbulent motions by subtracting the local mean velocity from the total velocity field, defined over a fixed spatial scale \citep[see e.g.,][]{Dolag_2005, Vazza_2009b}. Another commonly used approach is to compute velocity dispersion in  spherical shells and use it as proxy for level of turbulence \citep[e.g.][]{Lau_2009} for a given galaxy cluster. Even though these algorithms are conceptually simple and easy to implement, the fixed length scale does not capture the inherently multi-scale nature of turbulent processes.

\subsubsection{Reynold's Decomposition via multi-scale filtering}
\label{sec:methods_RD}
To address this problem, \cite{Vazza_2012,Vazza_2017} proposed the multi-scale filtering algorithm, which iteratively determines the spatial scale below which small-scale velocity fluctuations, i.e. turbulence, dominate over coherent motions. We indicate this as $ \mathcal{L} _{\rm{filter}}(\vec{x}) $, defined in any point of space $\vec{x}$. Such an approach allows us to extract the purely turbulent velocity field, $ \vec{v} _{\rm{Turb}} $ without fixing a spatial filtering length. Based on these works, we design a similar recursive algorithm to separate the mean (i.e. the bulk) velocity of a gas cell in the simulations from the turbulent velocity, given the total velocity field (i.e as predicted by simulations).

{The algorithm is iterative and is applied independently on each gas cell of the relevant simulation domain of interest, starting at its position $ \vec{x} = (x,y,z)$. Whereas we focus on the ICM at the centre of clusters (Section~\ref{sec:methods_defs}), the algorithm uses information from gas cells that are beyond the focus region, as it will be clear below.}

{The basic idea is that measurements of the velocity field within a spherical aperture of radius $ \mathcal{L}^{n}$ around any given gas cell are made at every iterative step $n$ and that such an aperture is increased at subsequent iterations until a convergence criterion is met.}

{For each relevant gas cell, at the first iteration, we take as initial scale $ \mathcal{L}^{n=0} = 2 r$, where $ r$ is the radius of the gas cell (whereby we approximate the Voronoi cells as spheres).

At each iteration $n$, three quantities are measured, all local at the position $ \vec{x}$ of the given target gas cell:}

\begin{enumerate}[leftmargin=1.5em]

  \item The bulk velocity, as
    \begin{equation}
      \vec{v} _{\rm{Bulk}, n} (\vec{x} | \mathcal{L} ^{\rm{n}}) = \frac{\displaystyle \sum_{i \in \mathcal{L} ^{\rm{n}}} w_{i}  \vec{v}_{i}}{\displaystyle\sum_{i \in \mathcal{L} ^{\rm{n}}} w_i}
    \end{equation}
    where the sum is over all gas cells within the spherical volume of radius $ \mathcal{L} ^{\rm{n}} $ centred on the gas cell of interest, and where $ w_i $ and $\vec{v}_i $ denote the weights (in this work gas mass) and velocity of each gas cell $ i $.\\

  \item The turbulent velocity, as
    \begin{equation}
      \vec{v} _{\rm{Turb}, n} (\vec{x} | \mathcal{L} ^{\rm{n}}) = \vec{v} _{\rm{Total}}(\vec{x}) - \vec{v} _{\rm{Bulk}, n} (\vec{x} | \mathcal{L} ^{\rm{n}})
    \end{equation}

  \item The convergence metric, the relative change in the turbulent velocity compared to the previous iteration:
    \begin{equation}
      \tau = \max_{i=x,y,z} \left| \frac{v^n_{i, \rm{Turb}} -  v^{n-1}_{i, \rm{Turb}}}{v^{n-1}_{i, \rm{Turb}}} \right| 
      \label{eq:rd_convergence}
    \end{equation}

\end{enumerate}

{If the latter is smaller than a given chosen tolerance, then the algorithm stops. Otherwise, a subsequent iteration is undertaken with an increased aperture, }

\begin{equation}
  \mathcal{L} ^{n+1}(\vec{x} ) = \max (\mathcal{L} ^n + r, [1 + \chi ] \mathcal{L} ^n)
  \label{eq:rd_increment}
\end{equation}
where $ r$ is the size of the gas cell of interest and $ \chi $ is a small number, here fixed to $ 0.05 $, to prevent slow convergence.

Our algorithm returns the filtering scale over which turbulent and bulk motions are decomposed in an adaptive manner, namely the local, cell-by-cell {based} turbulence filtering scale $ \mathcal{L} _{\rm filter} $, {that is $\mathcal{L} _{\rm filter} = \mathcal{L} ^{\rm m} $ with {\tt m} the iteration when the algorithm converges for the target gas cell. It hence returns the turbulent velocity $\vec{v}_{\rm{Turb, n}}(\vec{x} | \mathcal{L}_{\rm filter})$ for each relevant gas cells at position $ \vec{x}$, with filtering scale $ \mathcal{L} _{\rm{filter}}(\vec{x}) $.}

{In practice, $ \mathcal{L} _{\rm filter} $ for each gas cell represents the spatial scale below which small-scale velocity fluctuations dominate over the large-scale coherent motions. Vice-versa, for each gas cell $ \mathcal{L} _{\rm filter} $, represents the spatial scale above which bulk motions dominate, so that $ \mathcal{L} _{\rm filter} $ can also be interpreted as a bulk-flow coherence scale.}

Importantly, our algorithm involves three free parameters. First, the initial outer length scale of turbulence, $\mathcal{L}^{n=0}$, which we set to twice the local gas cell size. This choice is motivated by the adaptive nature of {\tt AREPO}, where cell sizes vary with gas density (with denser regions having higher resolution). Setting this value too large can lead to over-smoothing of bulk velocities, while setting it too small introduces noise due to low number of gas cells for reliable bulk velocity estimation. Second, the increment factor $\chi$ that controls how rapidly the outer length scale grows between iterations (Eq.~\ref{eq:rd_increment}). Setting $\chi$ too small increases convergence time, while setting it too large can cause over-smoothing of the bulk velocity field. After extensive testing, we find $\chi = 0.05$ offers an optimal balance between computational efficiency and accuracy. Third, a $ \tau_{\rm{tol}} $ sets the threshold for maximum relative change in turbulent velocity components between consecutive iterations (Eq.~\ref{eq:rd_convergence}), below which the algorithm converges.  In our tests, we find that setting $\tau_{\rm{tol}} = 0.1$ produces physically realistic turbulent velocity fields.
These parameters together determine the sensitivity of the separation between bulk and turbulent motions, {which we extensively characterize below and in Appendix~\ref{sec:methods_scales}}.

\begin{figure}
  \centering
  \includegraphics[width=1\linewidth]{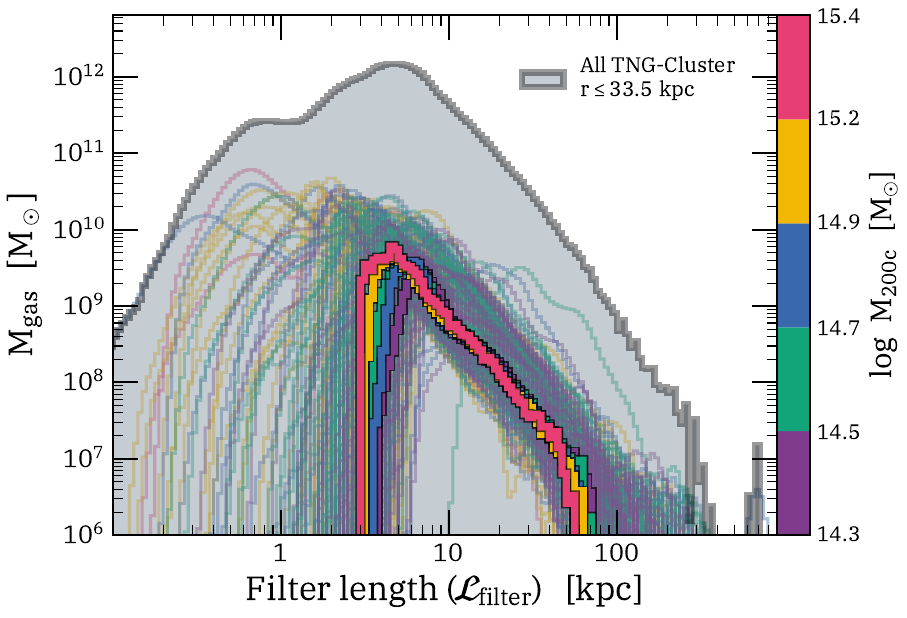}
  \caption{
    {{\bf Distribution of the filtering scale $\mathcal{L}_{\rm filter}$ that separates bulk vs. turbulent motions within the central regions of TNG-Cluster.} Thin lines indicate individual systems, whereas thick curves give the median results in cluster mass bins. $\mathcal{L}_{\rm filter}$ represents the local spatial scale above which bulk motion dominates over small-scale velocity fluctuations (i.e. turbulence). Based on the outcome of TNG-Cluster and the assumptions adopted in our algorithm, for the majority of the ICM mass in cluster cores and for the typical cluster, turbulence dominates on scales $\lesssim 5-10$ kpc.}
  }
  \label{fig:FilterLength}
\end{figure}

\begin{figure*}
  \centering
  \includegraphics[width=0.93\textwidth]{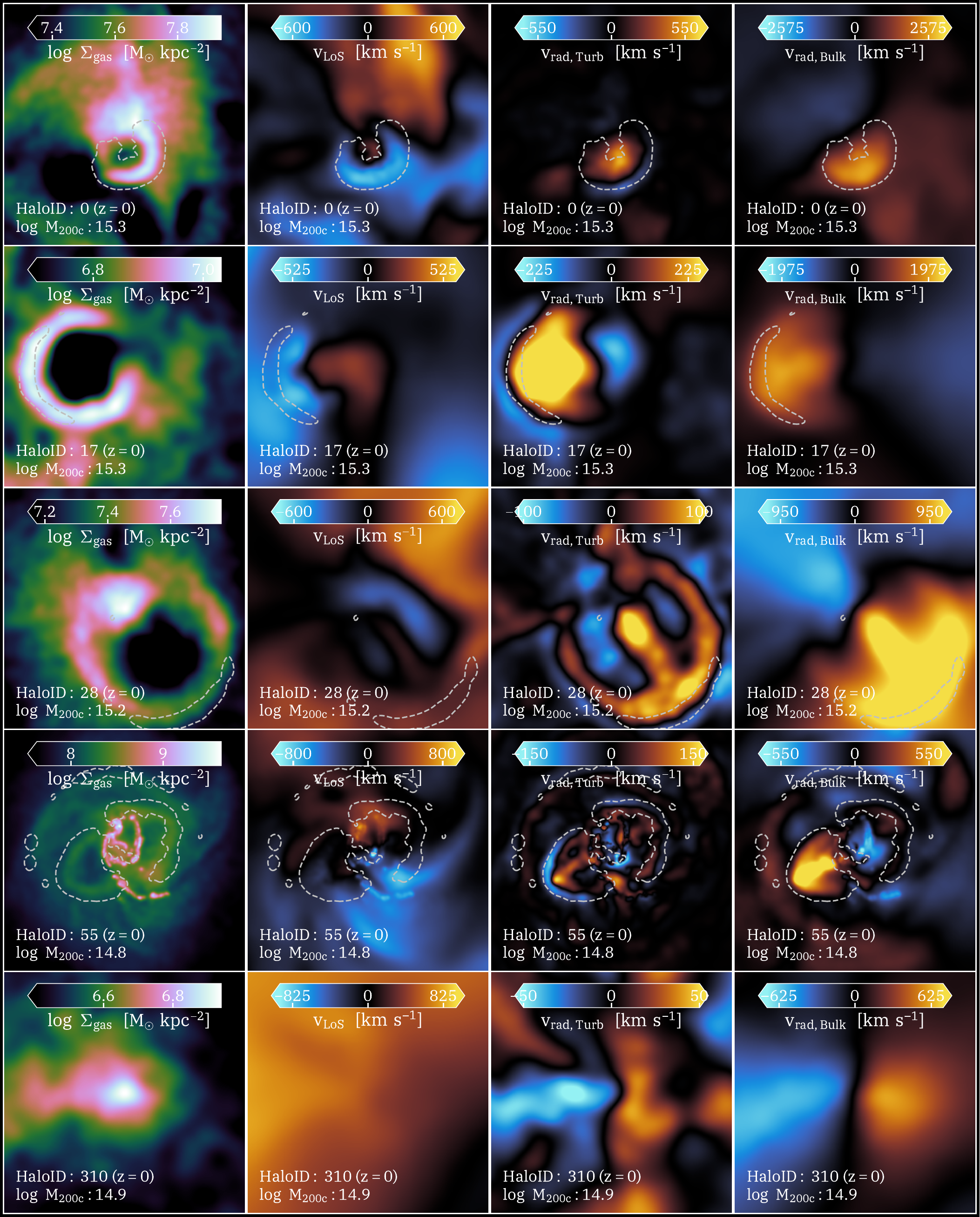}
  \caption{{\bf Spatially-resolved turbulent and bulk motions in the cluster centres for a subset of representative TNG-Cluster systems.}
    Each row shows a different cluster projected along the $z$ axis of the simulation box. The columns represent gas quantities, from {\bf left $ \to $  right}: projected density, projected line-of-sight (LoS) velocities, projected radial turbulent velocities, and projected radial bulk velocities. These are all within a central region of length $2\times33.5$ kpc, to mimic the extent of the FoV of a single {\it XRISM} pointing at the redshift of the Perseus cluster. White contours indicate, qualitatively, regions of rapid subsonic flow with local Mach numbers in the range 0.25 – 0.9. We plot these contours to visually highlight the expanding gas shells where radially-outward shocks and accompanying X-ray cavities are located \citep[see][for a detailed analysis of shocks and cavities in TNG-Cluster]{Prunier_2025c}. Bulk radial velocities de facto represent outflows, while turbulent radial velocities reveal downstream turbulence, which is strongest near shocks. One cluster (4th row) exhibits bipolar outflows with associated turbulence ($\sim 100$ km s$^{-1}$). The velocity field morphologies vary significantly from cluster to cluster, but turbulent velocities are consistently a small fraction of the total velocities -- a trend seen throughout the TNG-Cluster sample.
  }
  \label{fig:RD_collection}
\end{figure*}

\subsubsection{Functioning and spatial scales of our decomposition}
Fig.~\ref{fig:RD_Schematic} presents a schematic of our algorithm and the complex ICM velocity morphology in a TNG-Cluster system with a rising X-ray cavity at $z=0$. The top panel shows projected gas density (left) and projected total velocity (right) in a 200 kpc–thick slice centred on the cluster, with a field of view of $2\,\RFC$ on a side. The bottom panels zoom into the core: a 20 kpc–thick slice with a 67 kpc field of view (left), and a 10 kpc–thick slice with a 40 kpc field of view (right). Black contours highlight regions of fast subsonic flow (with local cell Mach numbers $\mathcal{M} = v/C_s \in [0.25, 0.9]$), serving as a visual guide to trace the expanding gas shells associated with a rising X-ray cavity and shock front. The bottom-left panel shows projected gas density, while the bottom-right panel shows line-of-sight (LoS) velocities for each Voronoi gas cell in the same slice. The multi-scale complexity represented in Fig.~\ref{fig:RD_Schematic} provides support to the adoption of a multi-scale filtering to separate bulk vs. turbulent motions.

In fact, Fig.~\ref{fig:RD_Schematic} shows that the ICM velocity field can be highly complex, with multiple coherent motions on different spatial scales. Our algorithm separates these coherent flows from turbulent motions by identifying, for each gas cell, the local {spatial} scale of coherence.  As an example, in the bottom-right panel we illustrate {how this scale is searched} for the central gas cell (purple star): the dashed circle indicates the initial scale $\mathcal{L}^{0}$, and the solid circle the final scale $\mathcal{L}_{\rm{filter}}$ {where our algorithm converges}.

Smaller values of $ \mathcal{L} _{\rm{filter}} $ (which also implies faster convergence) usually indicate that the gas cell is located in a region where the bulk motions dominate the gas velocities, whereas a larger value of $ \mathcal{L} _{\rm{filter}} $ usually indicates that the gas cell is located in a region where the turbulent motions dominate the gas velocities.

{Fig.~\ref{fig:FilterLength} shows the mass distribution of $\mathcal{L}_{\rm filter}$ for all gas across all the clusters in TNG-Cluster within the central regions. Each thin curve represents one cluster colour-coded by the mass-bin, while the thick curve represents the median distribution in the mass bin. On average, the coherent motions starts to dominate at scales of $\gtrsim 20$~kpc. Conversely, based on the outcome of TNG-Cluster and the assumptions adopted in our algorithm, for the majority of the ICM mass in cluster cores and for the typical cluster, turbulence dominates on scales $\lesssim 5-10$ kpc. These values can hence be taken as the typical filtering scales for the turbulence values quantified throughout this paper. However, we also note that there is a significant amount of gas mass for which bulk motions dominate on $ \mathcal{L} _{\rm{filter}} \lesssim 1 $ kpc, indicating that cluster cores have a very complex velocity field. }

We compare all results based on our adaptive multi-scale filtering with fixed scale filtering in Appendix~\ref{app:Comparison_FixedFilteringLength}.

\begin{figure*}
  \centering
  \includegraphics[width=1\linewidth]{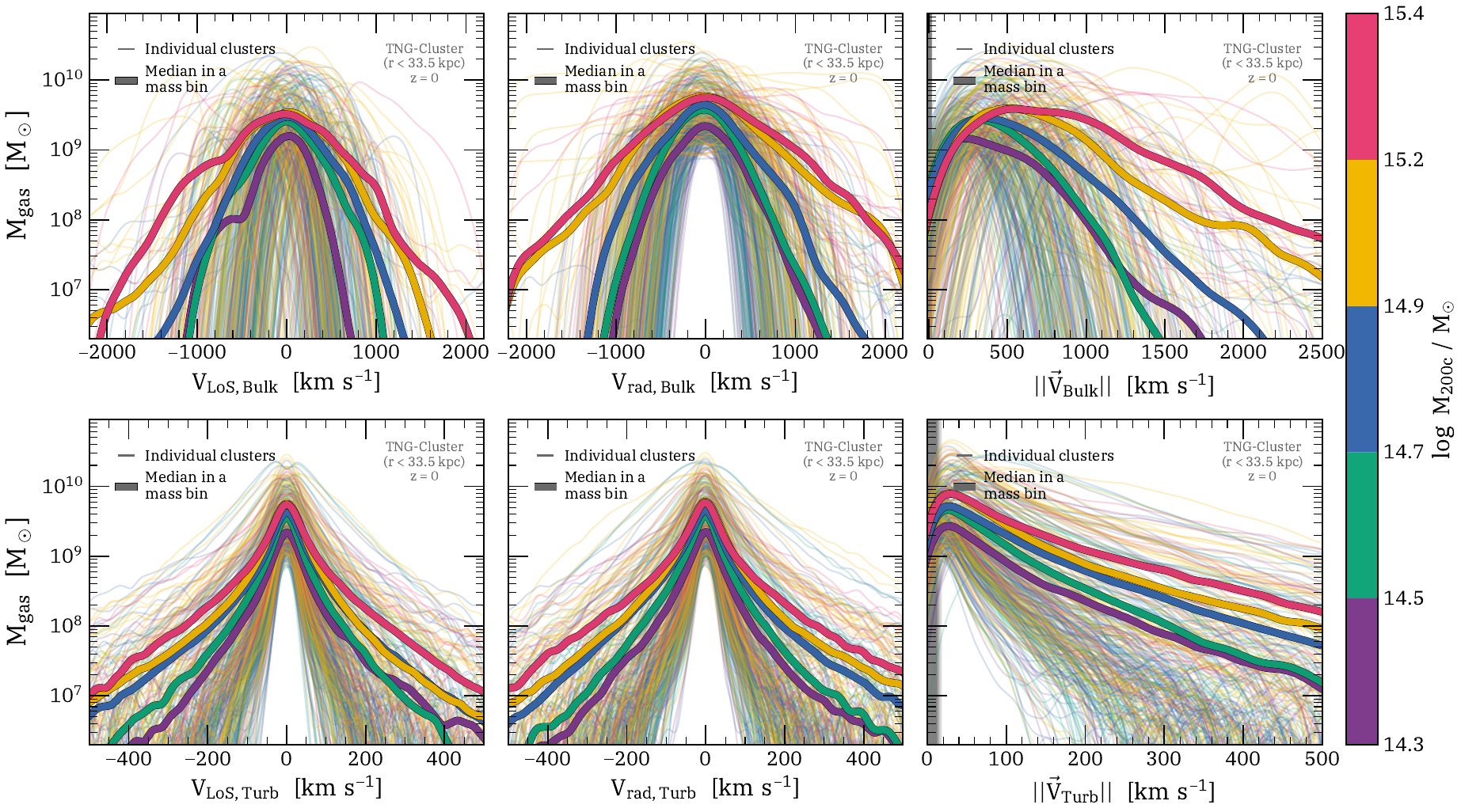}
  \caption{
    {\bf Velocity distributions of bulk and turbulent motions of the hot X-ray-emitting gas ($T > 10^{5.5}$ K) in the centres of clusters of the TNG-Cluster simulation.}
    {In all cases, we quantify the distributions by providing the amount of ICM gas in each velocity bin. From \bf Left} to {\bf right}, we show the distributions of line-of-sight (LoS), radial, and 3D velocities, all within the central $33.5$ kpc sphere of each cluster. The {\bf top} row shows the bulk velocity distributions and the {\bf bottom} row shows the turbulent velocity distributions. Thin curves denote individual clusters; bold curves represent the median velocity distribution in each cluster mass bin. Bulk velocity distributions are broad and strongly mass dependent, with bulk velocities reaching values of $\mathcal{O}(\pm 1000)$ km s$^{-1}$. On the other hand, turbulent velocity distributions are much narrower ($\mathcal{O}(\pm 100)$ km s$^{-1}$) and symmetric: {they exhibit somewhat weaker cluster-mass dependence than bulk motions but still more massive clusters clearly show }  broader and more extended high-velocity tails.
  }
  \label{fig:vel-distribution}
\end{figure*}

\section{Gas kinematics in TNG-Cluster's centres}
\label{sec:results}

The TNG-Cluster suite naturally returns a diverse population of galaxy clusters \citep{Nelson_2014}, each characterized by complex density and thermodynamics fields \citep[e.g.][]{Braspenning_2026, Chatzigiannakis_2026} and kinematics \citep{Ayromlou_2024, Truong_2024, Prunier_2025c, XRISM-Truong_2025}, in turn driven or shaped by a host of physical processes, from galactic feedback to gas accretion from the large-scale structure. Here we build upon previous works to showcase and quantify such diversity and complexity by explicitly focusing on the {\it true} turbulence and bulk components of the gas motions, separately.

\subsection{Spatially-resolved complexity and diversity of the ICM}
\label{sec:results_visuals}

Fig.~\ref{fig:RD_collection} shows the gas density and velocity fields of a subset of $z=0$ systems from TNG-Cluster and demonstrates the spatially-resolved complexity of turbulent motions in the ICM.

Each row represents one simulated cluster and the columns shows, from left to right: gas density, gas LoS velocities, gas radial turbulent velocities, and gas radial bulk velocities, in a central region with side length of $ 67 $ kpc, all projected along an arbitrary axis of the simulation's box ($z$ direction). We also over-plot the contours of regions characterized by Mach numbers (i.e. $v/C_s$) in the range $0.25-0.9$ (white contours). 
As a reminder, throughout this paper, we always refer to mass-weighted quantities, and we omit the specification from now on (Section~\ref{sec:methods_defs}).

Most of the clusters in Fig.~\ref{fig:RD_collection} exhibit radially-outward moving shocks, and these are often \citep[but not always, see][]{Prunier_2025c} associated to a cavity . In fact, while the LoS velocities may not always reveal the presence of cavities, the bulk radial velocities clearly show the outflows and the turbulent radial velocities nicely recover the down stream turbulence at the wake of the cavities, with the turbulent velocities being largest closest to the shocks. This pictures aligns well with theoretical expectations, whereby a large downstream turbulence is expected across shocks boundaries. Interestingly, one of the systems shows a bi-polar outflow, within which some level of turbulent motions is also present, $ \sim  100\rm{s ~km / s} $.

In addition to the examples of Fig.~\ref{fig:RD_collection}, we note that TNG-Cluster haloes have unique velocity field topologies, with turbulent velocities typically accounting only for a small fraction of the total ICM velocities -- this is not only the case for the systems shown here, but across a vast range of clusters simulated within TNG-Cluster.

\subsection{Broad velocity distributions}
\label{sec:results_distribs}

We quantify the statistical properties of the ICM kinematics by analysing the full velocity distributions of the gas in the TNG-Cluster cluster centres at $z=0$. Fig.~\ref{fig:vel-distribution} shows the distributions of the gas velocities for bulk (top row) and turbulent (bottom row) motions: from left to right, the LoS (here along the $z$ axis of the TNG-Cluster box) component, the radial component, and the velocity magnitude. In all cases, and as per design (Section~\ref{sec:methods_defs}), we consider only the gas within the central regions of simulated clusters and here we select only the hot X-ray-emitting gas (i.e. $ T > 10^{5.5}~\rm{K} $, Section~\ref{sec:methods_defs}). Thin curves indicate all the individual clusters from TNG-Cluster; thick curves show medians in bins of cluster mass.

The bulk-velocity components (top) exhibit broad distributions spanning $\mathcal{O}(\pm 1000)$ km s$^{-1}$, with more massive clusters showing systematically larger velocities. In contrast, the turbulent components (bottom) display much narrower distributions, with the majority of gas having velocities of $\mathcal{O}(\pm 100)$ km s$^{-1}$. Notably, while the bulk LoS velocities show asymmetric distributions that vary with cluster mass, the turbulent LoS velocities are symmetric around zero. Likewise, both the bulk and turbulent radial velocity distributions are symmetric around zero but the turbulent velocity distributions are significantly narrower than their bulk counterparts, albeit with small fractions of the ICM gas having turbulent velocities exceeding many hundreds of  km s$^{-1}$. The broad distributions of the bulk ICM motions are due to the presence of large-scale coherent flows such as infalls, outflows and mergers, which do not necessarily exhibit specific symmetries. On the other hand, the narrow and symmetric distributions of the turbulent motions indicate a more isotropic behaviour.

The distributions of the velocity magnitudes reveal a clear mass dependence for bulk motions. The peak of the median curves increases from about $200$~km s$^{-1}$ for the $10^{14.3-14.5}\MSUN$ clusters to $\sim700$~km s$^{-1}$ for the highest-mass clusters ($\gtrsim10^{15.2}\MSUN$). Concurrently, the spread of the distribution (akin to the full width at half maxima) expands approximately from $ 600 $ to $1500$~km s$^{-1}$. {Conversely, while the distribution $||\vec{v}_{\rm Turb}||$ peaks at similar values,} the simulation clearly predicts that more massive clusters are characterized by higher-velocity turbulence tails.

We briefly pause to examine the tails of the turbulent velocity distributions, as this will also be relevant later on. In Appendix~\ref{app:FunctionalForm_TurbVelDist}, we provide evidence that the LoS turbulent velocity distributions of TNG-Cluster systems are better described by Lorentzian (i.e. Cauchy) than by Gaussian distributions, indicating a greater importance of the tails.
Moreover, for our lowest-mass clusters {$(\MTWOC = 10^{14.3 - 14.5}~\MSUN)$}, less than $0.5$ per cent of the total gas mass exhibit turbulent velocities larger than $ 500 $ km s$^{-1}$, whereas for the most massive systems {$(\MTWOC > 10^{15.2}~\MSUN)$} this increases to $2-3 $ per cent. {Even though the latter is still a very small fraction of the total gas mass in the centres, these clusters have more than an order of magnitude more mass with non-negligible turbulent velocities, when compared to 1-dex lower-mass clusters} -- this is probably reflecting that gas is more affected by gravitational dynamics and/or feedback processes in higher-mass clusters.

\begin{figure*}
  \centering
  \includegraphics[width=1\linewidth]{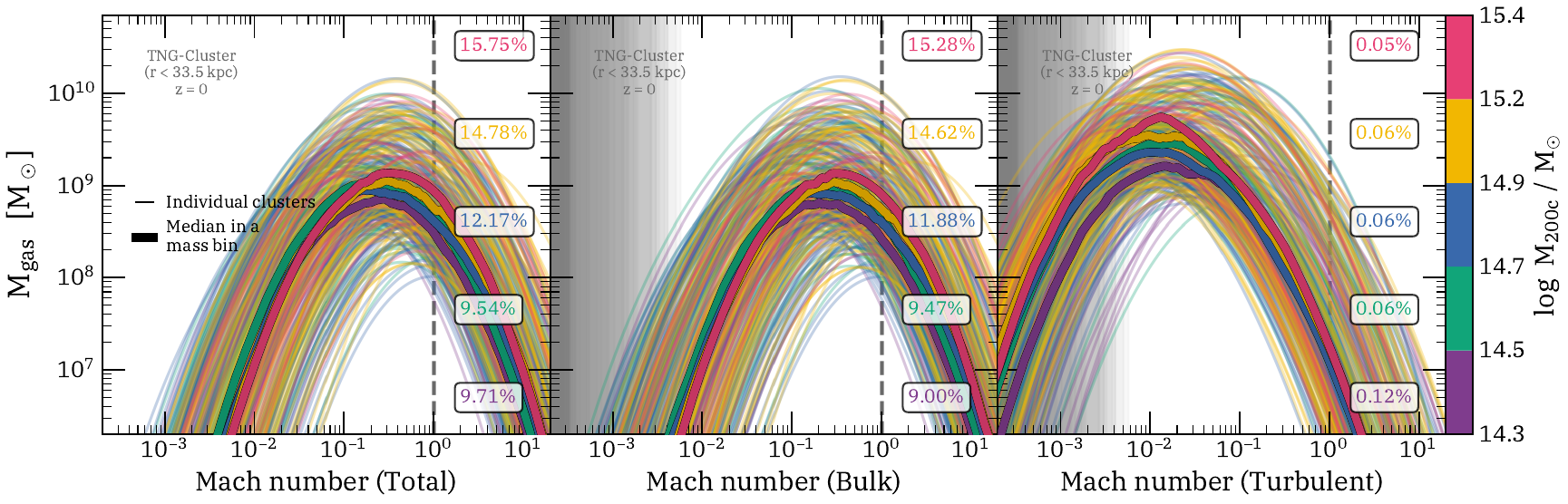}
  \caption{
    \textbf{Distributions of Mach numbers for the hot gas ($T > 10^{5.5}$ K) in the central regions of simulated clusters from the TNG-Cluster suite}.
    From left to right we show the distributions of Mach numbers ($\mathcal{M} = v/c_s$) measured from total, bulk, and turbulent velocities, respectively. Annotations and plotting scheme are as in Fig.~\ref{fig:vel-distribution}. The gray shaded region ($\mathcal{M} \lesssim 0.01$) marks the regime where our turbulent vs. bulk separation is not reliable. In most cluster centres, the largest fraction of ICM mass exhibit strongly-subsonic turbulence ($\mathcal{M}_{\rm Turb} < 0.2$), with turbulent Mach number distributions peaking at $0.01-0.03$ with little dependence on cluster mass. In contrast, bulk motions can extend to larger Mach numbers, with non-negligible fractions of ICM mass moving supersonically {(annotated by coloured numbers on each panel for each cluster mass bin)}, but with distributions peaking at Mach numbers of $0.2-0.5$.
  }
  \label{fig:vel-sound-distribution}
\end{figure*}

\subsection{Subsonic turbulence in cosmological systems}
\label{sec:results_machs}
To gauge the physical meaning and importance of the ICM velocity values, we transform them into Mach numbers ($\mathcal{M}$), i.e. the ratio of fluid velocity to the local sound speed. Fig.~\ref{fig:vel-sound-distribution}, presents the mass distribution of Mach numbers for hot gas ($T > 10^{5.5}~\rm{K}$) within the central sphere across all clusters in TNG-Cluster simulations for different velocity components. Annotations are as in Fig.~\ref{fig:vel-distribution}, with thin curves indicating individual clusters and thicker ones providing the cluster mass-bin medians. The gray shaded region at $ \mathcal{M} \lesssim 0.01 $ indicates the regime where our distinction between turbulent and bulk motions is not {effective}, given the functioning of the method (Section~\ref{sec:methods_RD}) and the choices of its free parameters (Section~\ref{sec:methods_scales}): it is based, for each cluster, on the 50th percentile of the ratio between turbulent velocity error ($\delta V_{\rm{Turb}}^{n, n-1}$, see Fig.~\ref{fig:RD_sensitivity}) and sound speed.

TNG-Cluster predicts that the majority of the hot gas in most of the cluster centres exhibits subsonic turbulence with $\mathcal{M}_{\rm{Turb}} < 0.2$ (90th percentile limit). Interestingly, the distributions of turbulent Mach numbers span a range broader by about 2 dex than those of the total (or bulk) motions. However, a much larger fraction of the ICM mass in cluster centres is associated with supersonic bulk motions (Mach number $>1$) than with supersonic turbulence, also on a cluster-by-cluster basis.

\begin{figure*}
  \centering
  \includegraphics[trim=0 0 0 60,clip,width=1\linewidth]{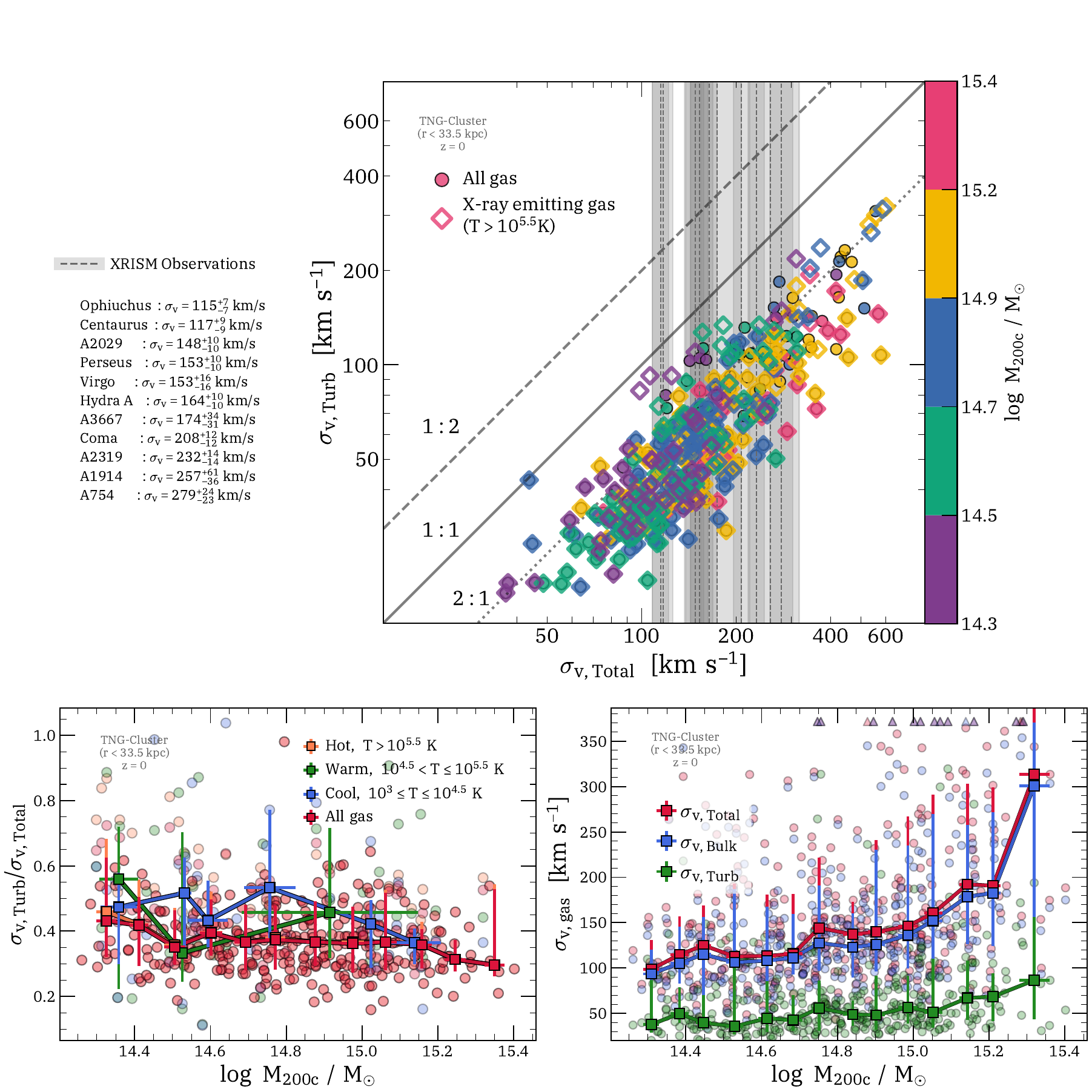}
  \caption{
    \textbf{Turbulent vs. total velocity dispersion as function of cluster mass and gas temperature in the central regions of TNG-Cluster systems at $z=0$.}
    In the {\bf Main panel}, we quantify the relation between 3D mass-weighted turbulent velocity dispersion ($\sigma_{v,\rm{Turb}}$) versus total velocity dispersion ($\sigma_{v,\rm{Total}}$), colour-coded by cluster mass $(\MTWOC)$ at $z=0$. Filled markers are based on all gas, while empty markers on hot X-ray-emitting gas only. The dashed lines indicate the 1:1, 1:2, and 2:1 relations. For context, we include observational constraints from recent {\it XRISM} observations as vertical gray bands. In the {\bf Bottom-left}, we show the turbulence-to-total fraction ($\sigma_{v,\rm{Turb}} / \sigma_{v,\rm{Total}}$) as a function of cluster mass for hot (red), warm (orange), and cool (blue) gas phases. Hot gas shows low turbulence fractions ($40-50$ per cent) with no clear mass dependence but significant cluster-to-cluster scatter; warm and cool phases exhibit higher turbulence fractions. In the {\bf Bottom-right} panel, the velocity dispersion for total (red), bulk (blue), and turbulent (green) components are shown as a function of cluster mass. Total  bulk, {and turbulent} velocity dispersions shows {similar} mass trend, {where the velocity dispersion increase by a factor $\sim 2$ between the lowest and the most massive bin.} In the bottom panels, one point represents one cluster; thick solid lines show median profiles with $16^{\rm{th}}-84^{\rm{th}}$ percentile error bars.
  }
  \label{fig:sigmaturb2tot}
\end{figure*}

\subsection{The large non-turbulent contributions to the velocity dispersion of the ICM}
\label{sec:results_sigmas}

Throughout the previous Sections we have seen that, according to TNG-Cluster, the turbulent velocities of the ICM are typically lower and reach lower extreme values than the bulk velocities. This is also reflected in the values of the velocity dispersions, as we discuss next.

We can indeed summarize the velocity distributions of Fig.~\ref{fig:vel-distribution} via the velocity dispersion. We measure the velocity dispersion of total, bulk, and turbulent velocities, separately, as per Equation~\ref{eq:sigma}, i.e. as mass-weighted second central moment of the respective velocity distributions. As we show in Appendix~\ref{app:FunctionalForm_TurbVelDist}, for the turbulent component  the velocity dispersion is a good proxy of the Full Width at Half Maximum (FWHM) of the Lorentzian curve that best fits the velocity distributions. It is also a quantity that is directly comparable to line-width measurements in X-ray spectroscopic observations.

To give an indication of the fractional contribution of turbulence to the total kinetic energy of a cluster in terms of velocity dispersions (or their ratio), a key assumption is implicit. If $ \vec{v} _{\rm{Total}}  = \vec{v}_{\rm{Bulk}} + \vec{v}_{\rm{Turb}} $, as is the case on a gas cell-by-cell {based} given our method, then:
\begin{align}
  {\sigma^2}_{v, {\rm{Total}}} &=  {\sigma^2}_{v, \rm{Bulk}} +  {\sigma^2} _{v, \rm{Turb}}\nonumber \\
  &\quad + 2{\rm{Covar}} \left( v_{\rm{Bulk}}, v_{\rm{Turb}} \right)
  \label{eq:sigmasums}
\end{align}
If the covariance term is negligible compared to the other two terms, then we can approximate {$\sigma^2_{v, \rm{Total}} \thickapprox \sigma^2_{v, \rm{Bulk}} + \sigma^2_{v, \rm{Turb}}$} and we can compare the turbulent velocity dispersion to the total one.

For the systems simulated with TNG-Cluster, it is indeed the case that the covariance term of Eq.~\ref{eq:sigmasums} is subdominant for the average cluster -- see Appendix~\ref{APP:Bulk_Turb_Covar} for a complete discussion. We can hence give an idea of the level of turbulence in a system by comparing {$\sigma _{v, \rm{Turb}}$} to {$\sigma _{v, \rm{Total}}$}. Importantly, this comparison allows us to gauge what fraction of chaotic motions (such as those captured by {$\sigma _{v, \rm{Total}}$}) are actually due to turbulence.

Fig.~\ref{fig:sigmaturb2tot}, main panel, shows the relation between the total ($\svtot$) and turbulent ($\svt$) 3D mass-weighted velocity dispersions for the central regions of galaxy clusters in the TNG-Cluster simulation at $z=0$. {The former is akin to what typically reported from observational analyses of X-ray spectra and so, for context, we include with vertical gray bands observational constraints from {\it XRISM} observations \mbox{\citep[measurements taken from][]{XRISM-Truong_2025}}. For the TNG-Cluster results, we} provide two markers per cluster, both colour-coded by cluster mass: one measured using all the gas in the FoV (filled) and one only selecting for the hot X-ray-emitting gas ($T > 10^{5.5}~\rm{K}$, empty markers). This figure shows that $ \sigma _{v, \rm{Turb}} $ is systematically lower than $ \sigma _{v, \rm{Total}} $, with the majority of clusters lying below the 50 per cent line: namely, according to TNG-Cluster, turbulence in cluster centres contributes less than half of the total velocity dispersion. As discussed in Appendix~\ref{app:Comparison_FixedFilteringLength}, this fractional contribution depends somewhat on the choice of the filtering scale for the measurement of turbulence, but in all studied scenarios the  turbulent contribution to the velocity dispersion never reaches unity for the average cluster.

The bottom-left panel of Fig.~\ref{fig:sigmaturb2tot} shows the turbulence fraction (\(\sigma_{v,\rm{Turb}} / \sigma_{v,\rm{Total}}\)) as a function of cluster mass for different gas phases -- hot, warm and cool, as per Section~\ref{sec:methods_defs}. One circle represents one cluster, while the solid curves with square markers represent the median values in a mass bin, with the errorbar representing the 16th-84th percentiles range, i.e. the cluster-to-cluster variation. The hot-gas case closely tracks the trend of the total gas because the hot phase dominates in mass (and volume) the gas in the ICM at $ z=0 $. In the hot phase, turbulence is the most subdominant, with \(\sigma_{v,\rm{Turb}} / \sigma_{v,\rm{Total}}\) of about 30-45 per cent (see also empty markers in the top panel). While the TNG-Cluster simulations do not predict a strong mass trend for the turbulence-to-total velocity dispersion fraction in any of the gas phases, it however exhibits a large cluster-to-cluster variation, by factors of up to $ 2-3 $. The simulations also show that the warm and cool gas phases exhibit higher (relative) levels of turbulence compared to the hot X-ray-emitting gas. This suggests that turbulent motions may be more prominent in the cooler gas, possibly due to enhanced gas dynamics in non-virialized regions.

The bottom-right panel of Fig.~\ref{fig:sigmaturb2tot} complements the main panel by showing the velocity dispersion for different dynamical phases (i.e total, bulk and turbulent) as a function of cluster mass. Each marker represents one cluster, while the the solid lines with square markers represent the median values in a cluster mass bin, as on the left. First, {$\svtot$ is almost exclusively set by bulk motions and is not sensitive to turbulence. This implies estimates of turbulence from $\svtot$ almost always over-estimate the true small scale fluctuations.} TNG-Cluster predicts larger velocity dispersions for more massive clusters for all the three dynamical components, with the trend appearing somewhat steeper for the Total and Bulk components -- ignoring the most massive cluster bin, $\sigma _{v, \rm{Total~or~Bulk}}$ increase from an average of about 100~ km~ s$^{-1}$ for $10^{14.3}\MSUN$ clusters to about 200 km s$^{-1}$ for $10^{15.2}\MSUN$ clusters, these numbers reading about 50 and 75 km s$^{-1}$ for the turbulence component.

\begin{figure*}
  \centering
  \includegraphics[width=0.95\linewidth]{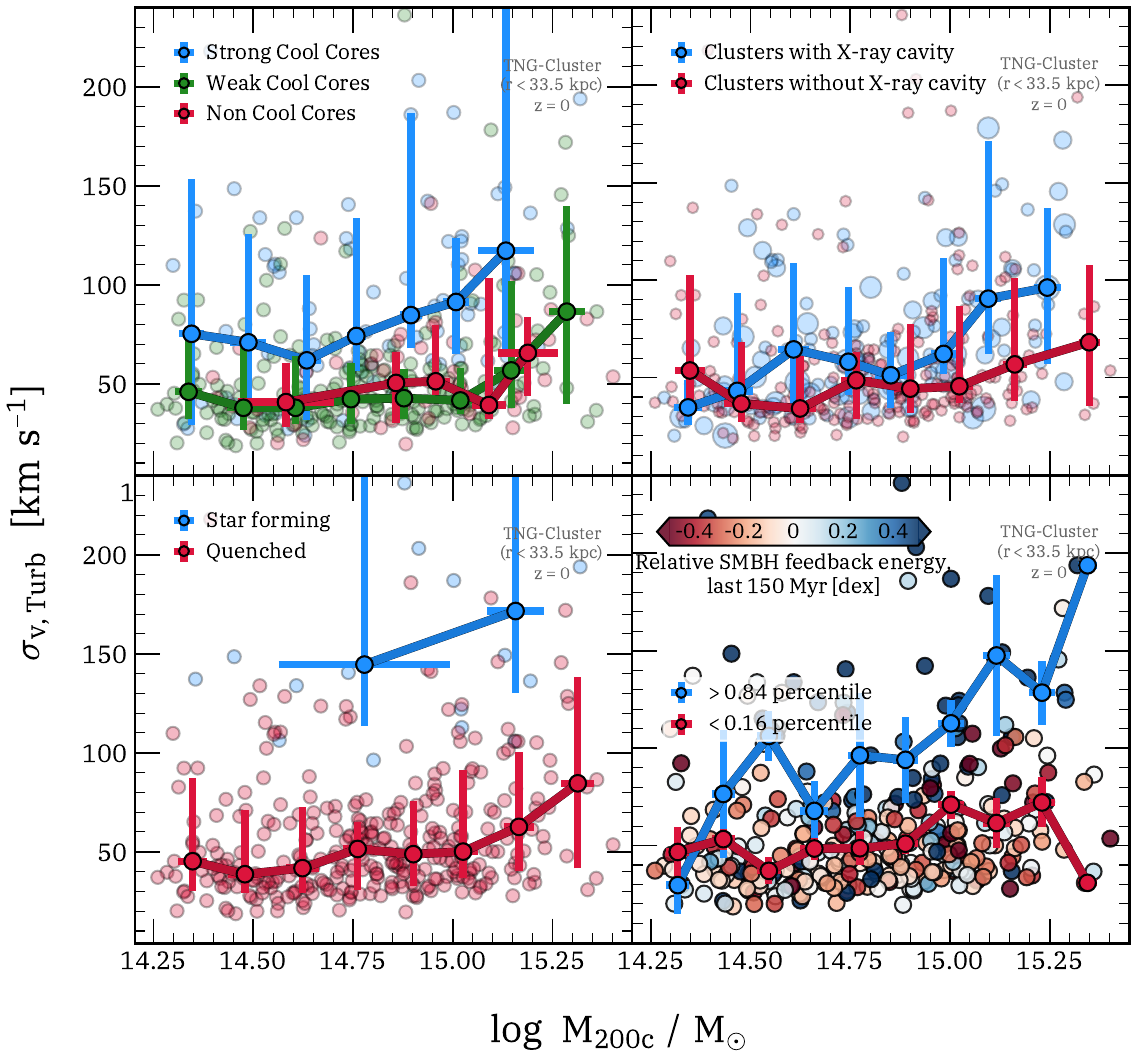}
  \caption{{\bf The dependence of the turbulent velocity dispersion on cluster properties related to SMBH feedback, according to TNG-Cluster.} In each panel we plot $\svt$ as a function of cluster mass ($M_{200c}$) in the cluster centres (central $33.5$ kpc sphere), with one marker representing one simulated cluster colour-coded by a cluster property and with solid curves representing median trends of specific clusters subsets (error bars for the 16th-84th percentile ranges). We focus on the hot X-ray-emitting gas ($T > 10^{5.5}$ K). Panels show properties primarily influenced by SMBH feedback: cool-core classification (SCCs=blue, WCCs=crimson, NCCs=green); presence of X-ray cavities (with cavities=blue, without=crimson; marker size indicates number of cavities); specific star formation rate of central galaxy (quenched vs. star-forming); and kinetic energy released as feedback by the SMBH at the center in the last 150 million years ({above and below the 84th and 16th percentiles for a given mass}). Strong cool-core clusters, clusters with X-ray cavities, clusters with star-forming BCGs and those with enhanced SMBH feedback at recent times show systematically higher turbulent velocity dispersions.}
  \label{fig:m-sigma_turb_SMBHProps}
\end{figure*}

\subsection{Drivers of cluster-to-cluster turbulence diversity}
\label{sec:results_correlations}

Although turbulence contributes less than bulk motions to the total velocity field (and $\sigma_{v,\rm turb} < \sigma_{v,\rm total}$; Section~\ref{sec:results_sigmas}), focusing on the turbulent component allows us to ask what drives the variation in ICM turbulence across clusters, including at fixed mass (Fig.~\ref{fig:sigmaturb2tot}).

{To investigate this, we systematically inspect a wide range of global and cluster-core properties to identify the strongest predictors of enhanced turbulent velocity dispersion. Interestingly, the properties that exhibited the strongest correlations—such as strong cool-core status, the specific star formation rate of the BCG, and the presence of X-ray cavities—share a common physical denominator: they are all intimately linked to the feedback from the central SMBHs. Conversely, large-scale dynamical parameters like merger history or relaxedness show surprisingly little correlation with core turbulence (a point we expand upon in Section~\ref{sec:discussion.mergers}). In Fig~\ref{fig:m-sigma_turb_SMBHProps} we show the strongest predictors, analysing how the turbulent velocity dispersion of the ICM depends on these properties}. In all panels of Fig~\ref{fig:m-sigma_turb_SMBHProps}, individual markers represent individual clusters, solid curves represent the median trends, and the error-bars indicate the 16th-84th percentile ranges (i.e. the cluster-to-cluster variations in bins of cluster mass).

\begin{enumerate}[label=(\alph*), leftmargin=2em, labelsep=0.75em, labelwidth=1.5em, listparindent=2em]

  \item \textbf{Cool coreness:} TNG-Cluster naturally returns a diversity of galaxy clusters in terms their central cooling times and entropies \citep{Lehle_2024, Lehle_2025}. Following the fiducial definition by \citet{Lehle_2024} based on central cooling time, we can classify the simulated clusters in Strong (SCCs, blue), Weak (WCCs, crimson) and Non Cool Cores (NCCs, green markers).

    Despite the considerable scatter and sometimes overlapping 16th–84th percentile ranges, SCC clusters are characterized on average by consistently higher turbulent velocity dispersions, $\svt $, than WCCs and NCCs (blue vs. green and red curves), by factors of $1.5-2.5$, i.e. by $40-60$ km s$^{-1}$. The SCCs also exhibit a much larger scatter across all the studied mass range, with tails of turbulent velocity dispersions exceeding $150-200$ km s$^{-1}$. The higher turbulent velocity dispersions in SCCs w.r.t to the WCCs and NCCs may stem from recent AGN activity, which is more common in SCC systems.

  \item \textbf{X-Ray Cavities :} TNG-Cluster naturally produces realistic X-ray cavities \citep{Prunier_2025b}, which we have shown to be driven by gas outflows in turn driven by the mechanical energy injections of the SMBHs \citep{Prunier_2025a}. We can hence distinguish between simulated clusters that exhibit identifiable X-ray cavities (blue markers; with marker size denoting the number of cavities per cluster, from 1 to 4) from those who do not (crimson markers), based on the X-ray surface brightness maps and classifications of \cite[][]{Prunier_2025a}.

    On average, clusters with cavities exhibit somewhat higher levels of turbulence compared to those without cavities, by up to a factor of 2 in certain mass bins. Beyond the averages (i.e. medians), cases of high turbulent velocity dispersions seem to be more often associated to multiple X-ray cavities per cluster. This enhancement might be due to two related processes: the SMBH feedback that drives the cavities also stirs the surrounding gas; and the cavities themselves act as buoyant structures that rise through the ICM, generating turbulence via hydrodynamic instabilities (see e.g., the second column, third row of Fig.~\ref{fig:RD_collection}). Additionally, the larger separation between the median lines for larger-mass clusters could be a result of more powerful SMBH feedback in more massive systems, which can create more cavities and drive stronger turbulence in the surrounding gas \citep[see][]{Prunier_2025a,Prunier_2025b}.

  \item \textbf{Specific Star Formation Rate (sSFR):} As shown by \cite{Nelson_2024}, the majority of BCGs in TNG-Cluster are quenched at $z=0$, but about 15 per cent of them are star-forming and many more have undergone periods of rejuvenation at recent times (\textcolor{blue}{Gottschewski et al. in prep.}). Here we report the sSFR for the BCG of each simulated cluster (i.e. the ratio of its star formation rate to its stellar mass), within twice its stellar half-mass radius, and considered it quenched if $ \mathrm{sSFR} < 10^{-11}~{\rm{yr^{-1}}} $.

    TNG-Cluster predicts that clusters with star-forming BCGs exhibit consistently and much higher turbulent velocity dispersions of their ICM compared to the quenched systems, by factors of a few. This correlation could be driven by a host of concurrent physical processes. In fact, we can at least point out that all star-forming BCGs in TNG-cluster are also SCCs, which show a similar preference for higher levels of turbulence.

  \item \textbf{Energy injected via SMBH feedback:} A very direct probe of the instantaneous or recent impact of SMBH feedback on the ICM is the amount of energy injected by the SMBHs in recent times. All TNG-Cluster systems at $z=0$ host low-accreting SMBHs at their centres \citep[e.g.][]{Prunier_2025a, Rohr_2024}, which, as per the IllustrisTNG model \citep{Weinberger_2017, Weinberger_2018, Pillepich_2021}, {release mechanical i.e kinetic feedback energy in discrete events i.e. not continuous in time}.  We hence measure the kinetic energy injected by the SMBH of each cluster over the last 150 Myr and separate clusters based on whether they have higher than average (blue) or lower than average (crimson markers) kinetic-mode feedback energy, in small bins of cluster mass. With solid curves, we also represent the median turbulent velocity dispersions of systems {exhibiting kinetic-mode feedback energy above the 84th (blue) and below the 16th (crimson) percentile levels}.

    Despite the considerable scatter, clusters that have undergone more than average SMBH feedback at recent times are characterized by consistently elevated turbulent velocity dispersions compared to those with less recent SMBH feedback, with a strong cluster mass dependence. As for the NCCs and clusters with X-ray cavities, we find larger variations among systems with more active and more substantial recent SMBH activity: this might be due to SMBH feedback being a highly stochastic process, which can create vastly-different systems, or to the possibility that clusters, albeit modulated by AGN activity, may be inspected at different times after the more recent feedback energy injections.
\end{enumerate}

The correlations of Fig.~\ref{fig:m-sigma_turb_SMBHProps} are all qualitatively consistent with each other and suggest a causal connection between SMBH feedback and the level of turbulence in the ICM of galaxy clusters. In fact, we have checked (but do not show) that these correlations also hold for the ratio of turbulent-to-total velocity dispersions. {These results are} in qualitative agreement with the findings of \cite{Truong_2024} based on TNG-Cluster, who also found higher ({with our definition}, total) mass-weighted velocity dispersions of Perseus-like clusters that host a more massive SMBH, a SMBH that is accreting at higher rates, or a SMBH that has released more kinetic feedback on its surroundings throughout its life time. Here we go a step forward and have shown with Fig.~\ref{fig:m-sigma_turb_SMBHProps} that clusters that are strong cool cores, exhibit X-ray cavities, and have undergone more SMBH feedback in recent times are characterized by larger {\it turbulent} velocity dispersions than the average cluster at the same mass. We investigate the causality of these correlations next.

\begin{figure*}
  \centering
  \includegraphics[width=\textwidth]{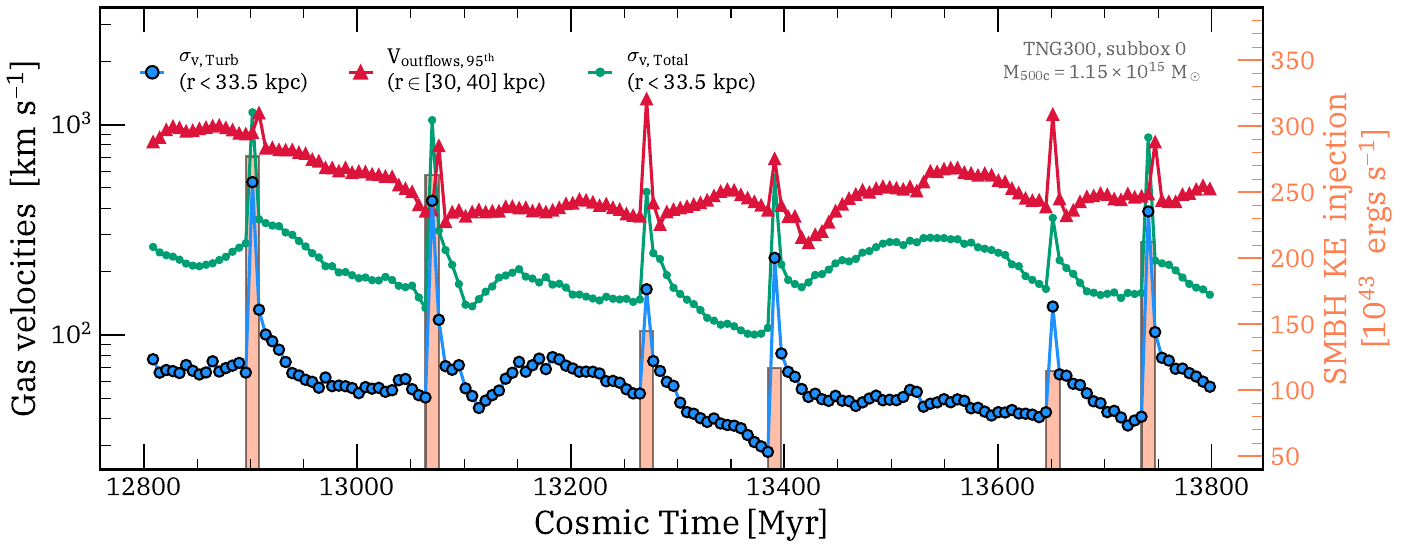}
  \caption{
    {\bf Time evolution of the gas kinematics in the central region of the most massive cluster in the TNG300 simulation, for which high temporal cadence output is available.}
    Coloured curves related to {\bf left $y$-axis} show total (green points) and turbulent (blue) velocity dispersion within 33.5 kpc from the centre, and the 95th percentile of the mass-weighted radial velocities in a shell at $30-40$ kpc (red, i.e. the high velocity outflows). Markers indicate the time of the actual snapshots (subbox-0 of TNG300). Orange vertical bars {\bf(right $y$-axis)} give the timings and approximate amounts of the mechanical energy injected by the central SMBHs. According to the IllustrisTNG model, episodes of SMBH feedback drive high-velocity outflows (up to about 1000 km~s$^{-1}$) and sharply enhance turbulent velocity dispersions (from $50-100$ km s$^{-1}$ to $200-500$ km s$^{-1}$), with a more pronounced effect on the turbulent rather than total velocity dispersion. Turbulence dissipates rapidly after feedback events, returning to pre-injection levels within $\sim 30 - 50$ million years.
  }
  \label{fig:TNG300_subbox0_times}
\end{figure*}

\subsection{SMBH feedback as a driver of turbulence in cluster centres}
\label{sec:results_timeevolution}

To further investigate the role of SMBH feedback in driving turbulence in cluster centres, we analyse subbox-0 of the TNG300 simulation \citep{Nelson_2019}, which, with an output cadence of about 8 Myr, enables exquisite sampling of the temporal evolution of turbulence in a cluster core.

The subboxes of IllustrisTNG are spatial cutouts from the full-volume runs, featuring a fixed comoving size and a much finer temporal output cadence than that of normal snapshots. subbox-0 of TNG300 is centred on the most massive cluster of the run at $z=0$, with a mass of $\MFC \sim 2 \times 10^{15} \MSUN$ and size of $ \RFC \sim 1610 $ kpc. Since TNG300 and TNG-Cluster employ the same physical model and numerical resolution, and because this halo falls within the TNG-Cluster mass range, it provides additional, consistent insight complementary to the TNG-Cluster sample. A similar use of the subboxes of IllustrisTNG to explore the temporal correlation between SMBH feedback events and their manifestations has previously been employed to understand the SMBH feedback-driven emergence and evolution of eROSITA-like bubbles in TNG50 galaxies \citep{Pillepich_2021}, as well as X-ray cavities \citep{Prunier_2025a} and shocks \citep{Prunier_2025c} in TNG-Cluster massive systems.

Here we apply the Reynolds Decomposition technique to each snapshot in the TNG300 subbox-0 between cosmic time, $ t = 12.3 - 13.3 $ Gyr and study the time evolution of bulk and turbulent velocities in the ICM core of the subbox-0 cluster in relation to the timings of SMBH feedback events.

In Fig.~\ref{fig:TNG300_subbox0_times}, coloured curves related to the left $ y- $axis represent the temporal evolution of total (green) and turbulent (cyan) velocity dispersion measured in a sphere of $ r = 33.5 $ kpc. These are contrasted to the radial outflow velocities in a spherical shell between $ 30-40 $ kpc from the centre, which represent bulk motions: of those, we provide the 95th percentile of the mass-weighted radial velocity values. The $x$-axis represents cosmic time in millions of years, with time progressing from left to right and $z=0$ being towards the right.  With the right $ y- $axis and with orange vertical columns we quantify the timing and amount of mechanical energy injected by the SMBH at the centre of the cluster -- since we don't store the exact time and amount of SMBH feedback in the simulation outputs, we approximate this by the gain in total kinetic energy of the gas between two consecutive snapshots in a central sphere of $ 40 $ kpc\footnote{While the gas in the ICM can move due to different physical processes, two of the main broadly accepted drivers are infall into the gravitational potential well and feedback from SMBHs. Changes in kinetic energy due to gas accretion occur on free-fall timescales, which are much longer than the times over which IllustrisTNG SMBHs inject their energy. Similarly, gas brought in by mergers would leave imprints on longer timescales. We can therefore confidently attribute changes in the kinetic energy of the gas in the central regions of subbox-0 to energy injection from SMBHs rather than large-scale structure effects.}.

As already established in studies of SMBH mechanical feedback in IllustrisTNG \citep{Nelson_2019TNG50, Zinger_2020, Pillepich_2021, Truong_2024, Prunier_2025a}, Fig.~\ref{fig:TNG300_subbox0_times} reiterates that feedback energy injection from the central SMBH drives very high-velocity outflows, reaching thousands of km s$^{-1}$. These appear as peaks in the temporal evolution of the outflow velocities (red curve). More novel and relevant for this paper is the time correlation between SMBH feedback events and turbulence. We find that after a kinetic feedback injection by the central SMBH, there is a strong enhancement in the turbulent velocity dispersion of the ICM (by factors of $4-10$), and this enhancement is more pronounced in the turbulent component than in the total velocity field. Our findings provide strong evidence that, in cluster centres, SMBH feedback is a key driver of turbulent gas motions.

Importantly, the simulation model predicts that this turbulence dissipates rapidly: the turbulent velocity dispersion typically returns to pre-injection values within $30-50$ million years after a feedback event.
{The} time evolution in Fig.~\ref{fig:TNG300_subbox0_times} suggests that, according to TNG300 and TNG-Cluster, feedback-driven turbulence in the centres of galaxy clusters is transient and requires frequent SMBH feedback energy injections (or duty cycles) to be sustained at high levels. {Similar conclusions were also drawn by \citet{Lehle_2025}, where the authors found individual feedback events (kinetic feedback) leave a very short impression on the gas, and the gas properties return to pre-feedback values shortly after}. This also explains the large cluster-to-cluster variations of Fig.~\ref{fig:m-sigma_turb_SMBHProps}, {as different cluster would have experienced feedback at different times resulting in different gas properties at the time of inspection,  $z=0$}.

\begin{figure*}
  \centering
  \includegraphics[width=1\linewidth]{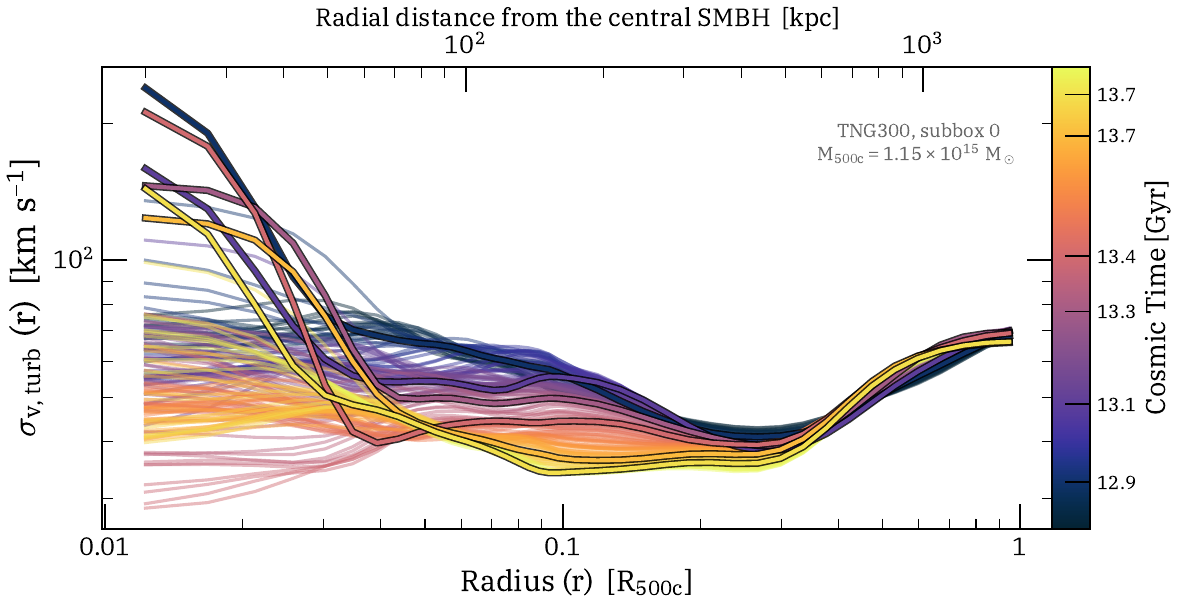}
  \caption{
    {\bf Time evolution of the radial profiles of turbulent velocity dispersion ($\svt$) in the most massive cluster of TNG300.}
    Colour shading from dark to light denote cosmic time from 12.3 to 13.3 billion years after the Big Bang, with each curve representing a snapshot with temporal cadence of about 8 Myr.
    Curves outlined in black correspond to epochs of SMBH feedback injection (as in Fig.~\ref{fig:TNG300_subbox0_times}, and as per black ticks in the colour bar). A characteristic `U-shaped' profile is in place at most times, but with very large variations in the central regions. SMBH-feedback episodes cause sharp, temporary enhancements in the turbulence of the core ICM, followed by rapid decay and outward propagation to larger radii. This may be interpreted as suggesting that, according to the IllustrisTNG model, SMBH feedback is responsible for driving turbulence on scales of $0.04-0.05\RFC$ ($60-80$ kpc from the centre), at least for {\em this particular cluster}, while turbulence in the outskirts may instead be sustained by quasi-steady cosmological gas accretion and merger-driven processes, at least over the 1 billion year time span considered here.
  }
  \label{fig:TNG300_subbox0_profiles}
\end{figure*}

Finally Fig.~\ref{fig:TNG300_subbox0_profiles} shows the time evolution of the 3D spherically-averaged radial profiles of the turbulent velocity dispersion of the same simulated galaxy cluster of Fig.~\ref{fig:TNG300_subbox0_times}. Different curves represent $\svt$ profiles at different cosmic times, colour-coded from dark to lighter shades from 12.3 to 13.3 billion years after the Big Bang. Curves highlighted by a black outline represents SMBH feedback events, i.e. the times when the SMBH injects mechanical energy (see also black horizontal ticks in the colour bar).


Qualitatively, a characteristic `U-shaped' trend manifest at most times, with larger values of $ \svt $ at smaller distances from the SMBH, followed by lower and then again higher levels of turbulence at intermediate and outskirts spatial scales, respectively. Importantly, at the times of SMBH feedback injection (black outlined profiles), there is a clear enhancement of turbulent velocity dispersion towards the centre, where it can reach a few hundreds km s$^{-1}$), and then a rapid fall by $ \sim 0.04-0.05 \RFC $ (or $ 60-80 $ kpc -- the latter can be interpreted as the initial injection scale of turbulence driven by SMBH feedback (and possibly via the creation of cavities and shocks), at least in the current implementation of the IllustrisTNG model and for this cluster. A few million years after a SMBH feedback event, the gradual decay in $ \svt $ in the central regions ($ \lesssim 0.05 \RFC $ ) and slight elevation between $ 0.05-0.2 \RFC $ may be a manifestation of turbulence cascading from the injection scale to smaller scales. This seems qualitatively in line with previous simulation works by \cite{Vazza_2011} who, using a fixed filtering length scale on galaxy cluster simulations, found that clusters move down in the $ E_{\rm{Turb}} -- E_{\rm{Thermal}} $ plane over a time span of $ 1-2 $ Gyr. In contrast, beyond $ 0.2 \RFC $, the turbulent velocity dispersion shows very weak time dependence. This suggests that, even though cosmic gas accretion and mergers may be the drivers of turbulence in the clusters outskirts, these processes may be in a quasi-steady state over Gyr timescales -- i.e there seems to be a constant injection of turbulence from these processes that maintains a steady level of turbulence in the outskirts.


\begin{figure*}
  \centering
  \includegraphics[width=1\linewidth]{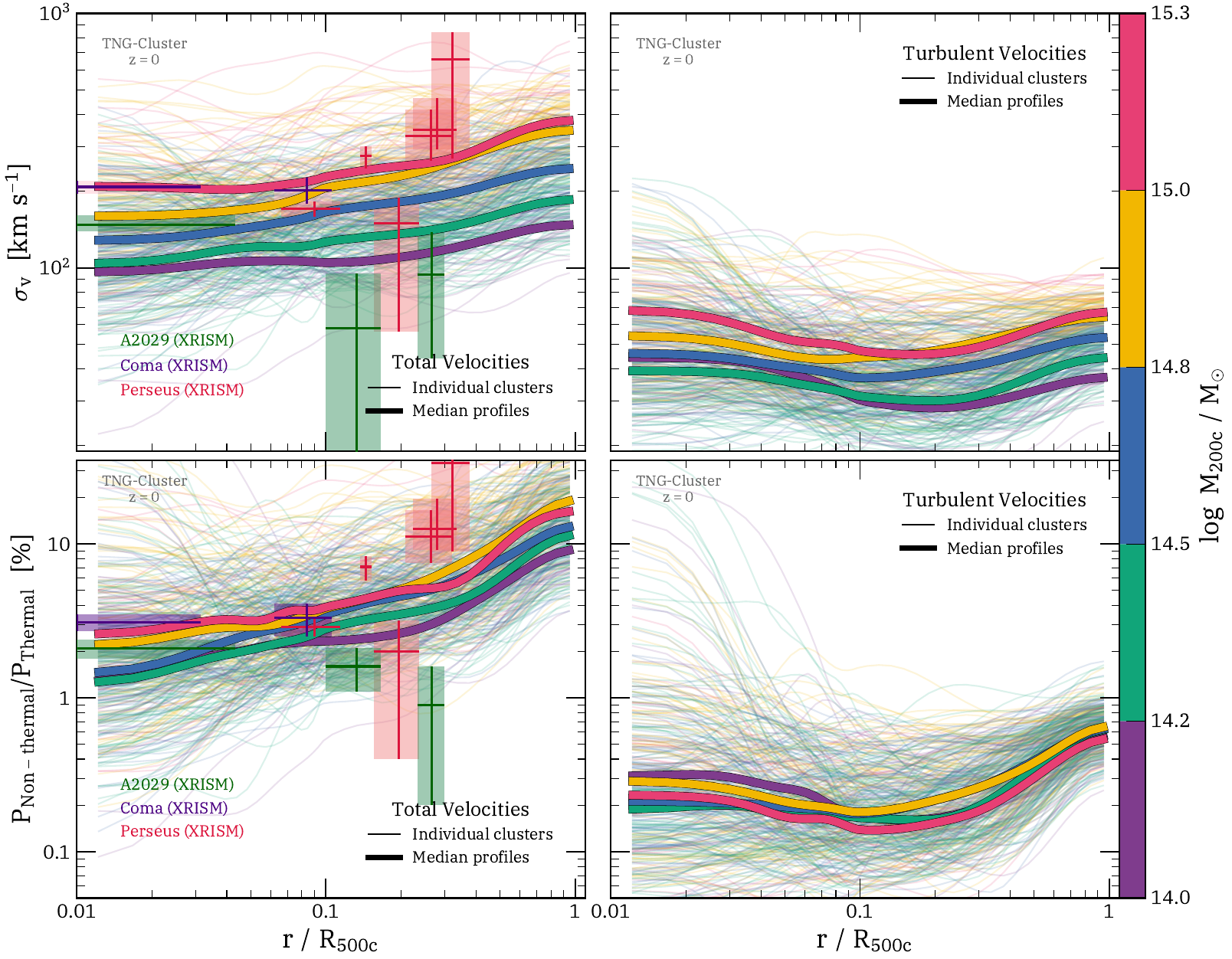}
  \caption{
    {\bf Radial profiles of gas velocity dispersion and non-thermal pressure support in TNG-Cluster systems.}
    In the {\bf Top panels}, we show the radial profiles of velocity dispersion for hot gas ($T > 10^{5.5}$ K), for total (left) and turbulent (right) velocities. Thin curves represent individual clusters; thick solid lines show the median profile in each mass bin.
    Scatter points mark {\it XRISM} observational measurements for Perseus, Coma, and Abell 2029. Total velocity dispersions increase with radius and cluster mass, reflecting stronger bulk motions in more massive systems and enhanced velocity dispersion in the outskirts due to accretion. Turbulent velocity dispersions exhibit a distinct `U-shaped' profile: elevated in cluster centres, dipping at intermediate radii ($\sim 0.2\,R_{500c}$), and rising again toward the outskirts, with only a weak mass dependence. In the {\bf Bottom panels} we show the radial profiles of the ratio of non-thermal (turbulent) to thermal pressure support, $P_{\rm{nt}}/P_{\rm{th}}$, estimated using total (left) and turbulent (right) velocity dispersions. While the total velocity-based ratio shows a strong mass trend and increases from centre to outskirts, the turbulence-only ratio remains low ($\lesssim 1$ per cent) and flat in the core, rising only beyond $0.2\,R_{500c}$, and shows no significant mass dependence. This highlights that true turbulent pressure support in cluster centres is minimal, consistent with recent X-ray observational constraints, and that bulk motions dominate the non-thermal pressure budget outside the core.
  }
  \label{fig:radial-prof.sigmas}
\end{figure*}

\section{Discussion and implications}
\label{sec:disc}
TNG-Cluster suggests that the turbulence in the central regions of galaxy clusters may well be driven, or at least temporarily enhanced, by AGN mechanical feedback, which concurrently drives high-velocity outflows of small fractions of ICM mass \citep[for SMBH-driven outflows see also][in the broader context of the IllustrisTNG model]{Nelson_2019TNG50, Pillepich_2021, Truong_2024, Prunier_2025a}. Importantly, typically -- i.e. for the average $z=0$ cluster and for the majority of ICM mass --, turbulent motions are largely subdominant {within the inner $33.5$ kpc. We deliberately chose this physical aperture for our core analysis to mimic the field of view of a single {\it XRISM} pointing at the redshift of the Perseus cluster, providing a controlled baseline to explore these dynamics.} In the following, we discuss how the findings of the previous Sections generalize and what their implications are for interpreting observational cluster data. The physics of the coupling between SMBH feedback and ICM turbulence, and its possible mediation via cavities and shocks, will be studied in future works, whereas the manifestations and drivers of turbulence in the circum-galactic media of Milky Way-like galaxies in the IllustrisTNG framework are quantified by \textcolor{blue}{Biba \& Nelson in prep.}.

\subsection{ICM turbulence beyond cluster centres}
\label{sec:disc_profiles}
We extend our analysis beyond the cluster centres (Sections~\ref{sec:results_distribs}--\ref{sec:results_timeevolution}) and beyond the case of one single cluster (Section~\ref{sec:results_timeevolution}, Fig.~\ref{fig:TNG300_subbox0_profiles}) to investigate how bulk and turbulent motions change as a function of cluster-centric distance, i.e. radially towards the cluster outskirts.

In Fig.~\ref{fig:radial-prof.sigmas} we show the 3D spherically-averaged profiles in bins of cluster-centric distance (normalized by the clusters' $\RFC$) of velocity dispersions (top) and of the non-thermal-to-thermal pressure ratios  (bottom, as per Eq.~\ref{eq:P_ratios}) for all TNG-Cluster systems at $z=0$, contrasting total vs. turbulent-only components (left vs. right). In all panels, thin (thick) curves denote individual clusters (mass-bin medians) and we only consider hot, X-ray emitting gas.
Finally, the individual markers and shaded areas denote observational measurements from {\it XRISM} for the Perseus (red), Coma (violet), and Abel 2029 (green) clusters \citep{XRISM_Perseus_2025, XRISM_A2029_2025, XRISM_Coma_2025}. These are shown for context and reference only, i.e. not for direct comparison, and are deliberately and consciously placed in panels where the simulated kinematic profiles are derived from the total (rather than turbulent-only) velocity field.

\subsubsection{`U-shaped' radial profiles of turbulent motions}
\label{sec:disc_profiles_Ushapes}

A key overarching take-home message from Fig.~\ref{fig:radial-prof.sigmas} is the large cluster-to-cluster variation in the kinematic profiles, already seen above in central quantities and for an individual cluster across cosmic times, but here manifest also at intermediate cluster-centric distances, i.e. at $0.1-0.5~\RFC$. This cautions against any strong conclusion from the comparison of single observed clusters to simulated individual clusters or median trends.

However, kinematic profiles averaged across many clusters \citep[see also][]{Ayromlou_2024} appear to be driven by, and may therefore provide insight into, the dominant underlying physical processes.

Comparing the top left to the top right panel of Fig.~\ref{fig:radial-prof.sigmas}, the velocity dispersions of the total gas velocities are, on average, larger than those of the turbulent component across entire clusters and not just in the centres (as demonstrated in Section~\ref{sec:results_sigmas}). Moreover, amid the large scatter, whereas the total velocity dispersion tends to monotonically increase towards cluster outskirts (with a clear cluster mass dependence), TNG-Cluster predicts a distinct (albeit not-too-pronounced) `U-shaped' trend for the median profiles of the turbulent velocity dispersion. Namely, $ \svt $ is elevated in the cluster centres, typically decreases to a minimum at around $ 0.1-0.2~\RFC $, depending on cluster mass, and then rises back up to central values towards the outskirts. For individual clusters (thin curves), the largest values of turbulent velocity dispersion are found in the central regions. This generalizes the findings of Section~\ref{sec:results_timeevolution} based on a single cluster at various evolutionary stages and suggest that `U-shaped' radial profiles of the turbulent velocity dispersion are common.

The monotonic increase of the total ICM velocity dispersion towards cluster outskirts is consistent with previous theoretical work \citep[see e.g.,][]{Lau_2009,Lau_2013,Vazza_2011}, according to which ongoing accretion of matter from the cosmic web increases the high-velocity tails of the distribution, thereby enhancing the velocity dispersion. In fact, more massive clusters have deeper gravitational potential wells that produce stronger gas motions, naturally leading to higher velocity dispersions and higher overall velocities \citep[see e.g.,][]{Munari_2013}.

The qualitative difference in the shape of the radial profiles of $ \svtot $ and $ \svt $ suggests that the physical drivers of turbulent motions are different from those of the bulk motions. As already discussed for the single cluster of Fig.~\ref{fig:TNG300_subbox0_profiles}, the `U-shaped' radial trend of the turbulent velocities is highly suggestive of a transition between two dominant physical drivers of turbulence \citep[e.g.][]{Vazza_2017,Simionescu_2019}. The central increase is likely driven by internal processes such as SMBH feedback and core sloshing, while the rise in the outskirts is driven by external accretion and substructure infall. The intermediate `dip' marks the region where the gas is most relaxed, being sufficiently distant from the central engine yet shielded from the active accretion at the virial boundary.

\subsubsection{Sub-percent to percent turbulent pressure support}
\label{sec:disc_profiles_Pratio}
In the bottom panels of Fig.~\ref{fig:radial-prof.sigmas}, we show the radial profiles of the ratio between non-thermal and thermal pressure, $ P_{\rm{nt}} / P_{\rm{th}} $, using the total kinetic energy (left) and isolating just the turbulent motions (right).

The non-thermal-to-thermal pressure ratio (Eq.~\ref{eq:P_ratios}) estimated using the \textit{total} velocity dispersion (bottom left) corresponds to the quantity typically accessible in observations. In addition to large cluster-to-cluster variations, TNG-Cluster predicts both a strong mass dependence and an increasing trend from cluster centres to outskirts. At the very centre, the median ratio ranges from one to a few per cent from the lowest- to the highest-mass clusters, while at the outskirts the median $ P_{\rm{nt}} / P_{\rm{th}} $ is in the $8-20$ per cent range. This radial increase is consistent with previous simulation work \citep[e.g.,][]{Lau_2009,Nelson_2014,Vazza_2018}, which attribute the enhanced non-thermal pressure in the outskirts to ongoing accretion and mergers, and is in line with the radial trend seen for the total velocity dispersion in the top left panel.

The pressure support from {\it {solely}} the turbulent component (bottom right) is consistently lower than the figures above, with little cluster-mass dependence and a weak `U-shape' radial dependence. Namely, the turbulent pressure support once bulk motions are removed is typically below the percent level across the full extent of the clusters, even though individual systems can exhibit $ P_{\rm{nt}} / P_{\rm{th}} $ ratios of up to $20-30$ per cent in their cores.


\begin{figure*}
  \centering
  \includegraphics[width=0.95\linewidth]{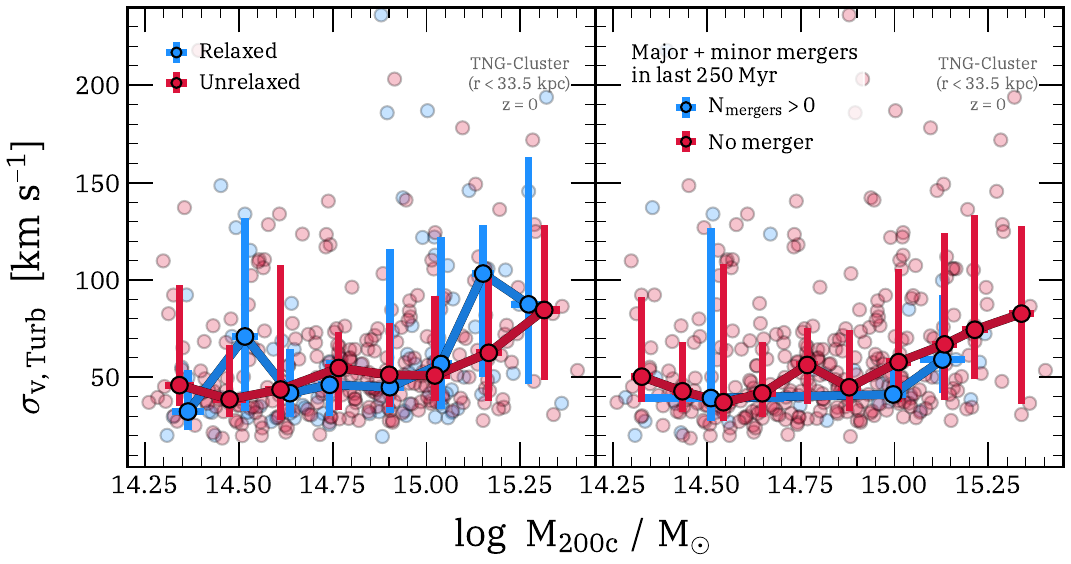}
  \caption{
    {\bf The apparent lack of dependence of the turbulent velocity dispersion in cluster centres on large-scale external processes.}
  Setup and annotations are as in Fig.~\ref{fig:m-sigma_turb_SMBHProps} and $ \svt $ is measured within a central $33.5$ kpc sphere around each cluster, for hot X-ray-emitting gas ($T > 10^{5.5}$ K). TNG-Cluster systems are separated based on  dynamical relaxedness (left) and whether they have had a major or minor merger in the last 250 million years (right). See text for details. According to TNG-Cluster, the recent merger or accretion history, or a current proxy thereof, does not leave a clear imprint on the levels of ICM turbulence in cluster centres. }
  \label{fig:m-sigma_turb_mergers}
\end{figure*}

\subsection{On other possible drivers of turbulence in cluster centres}\label{sec:discussion.mergers}

When interpreting the shapes of the radial profiles of turbulent velocity dispersion and pressure support above, we have repeatedly invoked cosmological gas accretion and mergers as drivers of enhanced motions in cluster outskirts. What about their effects on the centres? While the previous Sections clearly demonstrate that SMBH mechanical feedback can drive turbulence in cluster cores, we have not necessarily shown that this is the only or dominant mechanism. Indeed, many observed galaxy clusters exhibit patterns in central X-ray surface brightness maps that can be clearly associated with merging and accretion events \citep[see for e.g.][]{Walker_2017,XRISM-Perseus-extended_2026}.

Following the logic and annotations of Fig.~\ref{fig:m-sigma_turb_SMBHProps}, in Fig.~\ref{fig:m-sigma_turb_mergers} we plot the turbulent velocity dispersion in cluster centres as a function of cluster mass by distinguishing between clusters with or without a recent merger or accretion history.

\begin{enumerate}

  \item \textbf{Relaxedness:} We can classify TNG-Cluster systems into relaxed or unrelaxed based on {the criteria of \citet{Ayromlou_2024}. Namely, a cluster is considered relaxed if the separation between the halo's centre of mass and its most bound particle is $< 0.1 \RTWOC$ and if the mass fraction of the central subhalo relative to the total halo mass exceeds 0.85.}

    TNG-Cluster predicts no distinguishable difference among the two populations of clusters based on their level of gas-based relaxedness -- both relaxed and unrelaxed systems shows similar median trends and overlapping 16th–84th percentile ranges across the entire mass range. The only difference the simulation predict is that relaxed systems exhibit a slightly larger scatter in $ \sigma _{v, \rm{Turb}} $ compared to unrelaxed systems.

  \item \textbf{Mergers:} We can also distinguish between clusters that have experienced either a minor or major merger within the last \(250\) Myr (blue markers) and those that have not undergone any merger in this time span (crimson markers). The merger history is derived from the merger trees of the primary subhalo, with a major merger defined as one with a stellar mass ratio greater than \(1/4\) and a minor merger defined as one with a stellar mass ratio between \(1/10\) and \(1/4\).

    We find no trend in the $ \svt $  between systems with and without recent mergers.
\end{enumerate}

We have explored the possible dependence of the turbulent velocity dispersion in cluster centres on alternative proxies of recent merger and accretion history, finding no clear population-wide trends. However, the absence of such correlations does not necessarily imply a lack of physical impact of merging systems on turbulence in cluster cores, especially in a localized sense. Indeed, previous studies have reported clear connections between ICM turbulence and phenomena such as sloshing, which may be induced by mergers \cite[e.g.][]{ZuHone_2012,Mirakhor_2023}. Further work in this direction, particularly in the context of cosmological clusters including SMBH feedback, is therefore warranted.

\subsection{Implications for interpreting observational results}
\label{disc:comparison_XRISM}
Our analysis and results based on the outcome of the TNG-Cluster cosmological suite of galaxy clusters simulations offer a theoretical baseline for interpreting the high-resolution X-ray spectroscopy of the ICM currently being delivered by {\it XRISM}, and expected from future missions like {\it NewAthena}. Recent results from {\it XRISM} have revealed somewhat-surprisingly narrow emission lines in the cores of clusters such as Perseus, Coma, or Abell 2319, implying low levels of non-thermal broadening \citep{XRISM_A2319_2025, XRISM_Perseus_2025,XRISM_A2029_2025, XRISM_Coma_2025}. Our findings contextualize these observations in several ways.

First, our results (Sections~\ref{sec:results_sigmas} and \ref{sec:disc_profiles}) caution against interpreting the total line broadening strictly as turbulence, and calling the so-inferred gas velocity dispersion (which is by construction {\it total}, in our jargon) as turbulent velocity dispersion. According to our analysis of TNG-Cluster, the turbulent velocity dispersion ($\svt$) in cluster centres accounts for only a fraction of the total velocity dispersion ($\svtot$): $40-50$ per cent, in the average cluster and in our fiducial analysis (see also Appendix~\ref{app:Comparison_FixedFilteringLength}). The remainder is dominated by bulk, coherent motions  -- such as sloshing, inflows, and large-scale circulation driven by the hierarchical growth of clusters as well as possible diverse outflows from the AGN along the line of sight. This implies that observational estimates of turbulence derived solely from line broadening (without spatially resolving the velocity field to subtract bulk flows) likely overestimate the  turbulent energy density by a factor of $\sim 4$ (since $E \propto \sigma^2$).

Second, the distinction and quantitative difference between total and turbulent velocity dispersion has significant consequences for estimating the hydrostatic mass bias. Commonly it is thought that the non thermal pressure required to explain the hydrostatic mass bias of galaxy clusters is present in the form of turbulence. However, our analysis (Section~\ref{sec:disc_profiles_Pratio}) shows that while the ``total'' non-thermal pressure support in the core can reach a few percent for the average cluster, the support from \textit{true} small-scale turbulence is negligible ($\lesssim 1$ per cent). This indicates that turbulence is not the primary source of hydrostatic-equilibrium bias in cluster cores; so any significant deviations from equilibrium in these regions are likely driven by bulk motions or geometric asymmetries.

Third, the cluster-to-cluster variations predicted by TNG-Cluster, whether in central measurements (Sections~\ref{sec:results_sigmas} and \ref{sec:results_correlations}) or radial profiles (Section~\ref{sec:disc_profiles}) are large. Total and turbulent velocity dispersions span values across orders of magnitude both in cluster centres and at larger radii, depending on the cluster (and only partially on its mass). This highlights the need for caution when drawing strong conclusions from comparisons between individual observed clusters and single simulated systems or median trends.

Finally, our work suggests that the level of turbulence, albeit small, may be a sensitive probe of SMBH feedback physics. We find that systems with strong cool cores and X-ray cavities exhibit systematically higher turbulence than non-cool-core systems. This implies that future observations targeting a diverse sample of clusters could use the scatter in velocity dispersion measurements to constrain the duty cycles and coupling efficiencies of SMBH feedback models. The characteristic `U-shaped' radial profile of turbulence further suggests that spatially resolved spectroscopy (e.g., with IFUs on New Athena) could disentangle the contributions of AGN-driven turbulence in the core from accretion-driven turbulence in the outskirts. However, a lot of the information is encoded in the full distribution of ICM velocities, as we highlight in the next Section.

\begin{figure*}
  \centering
  \includegraphics[width=1\linewidth]{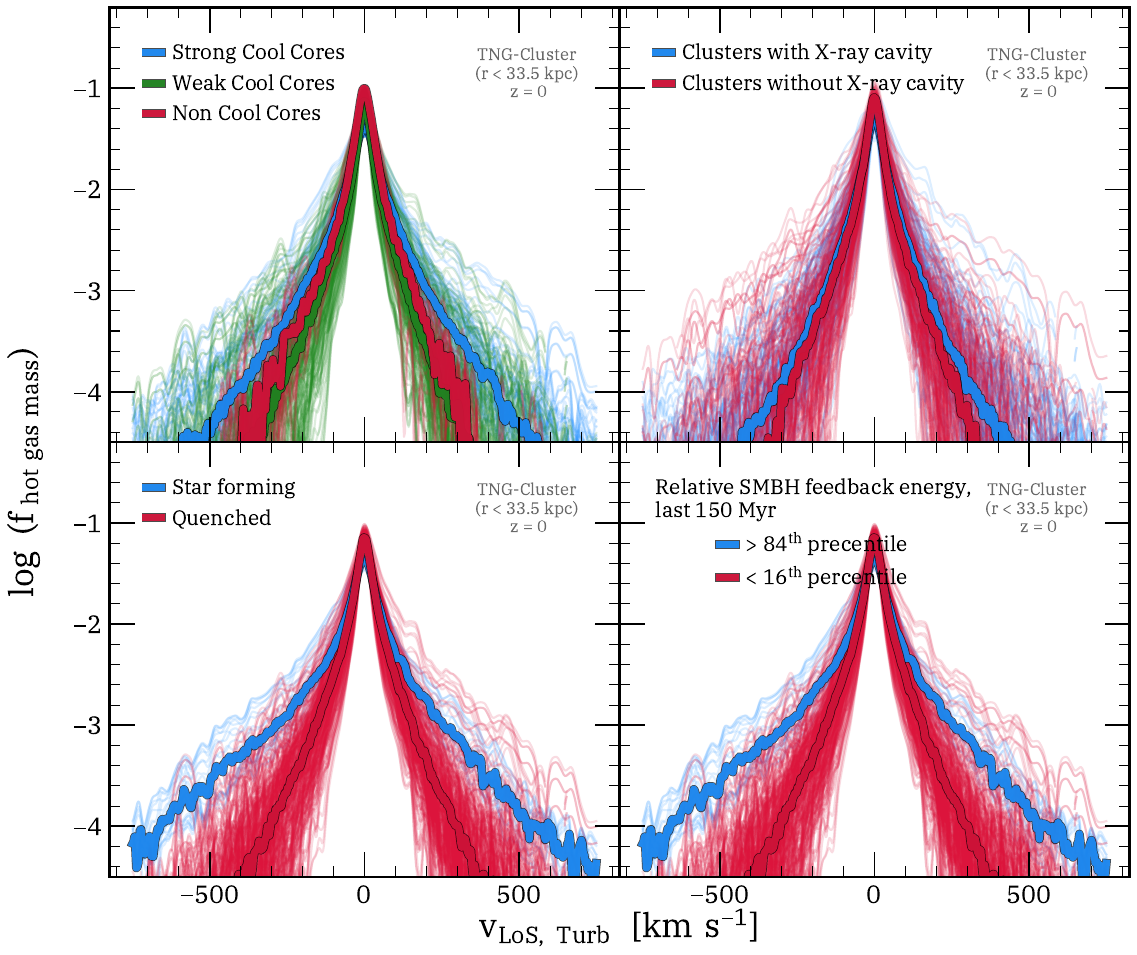}
  \caption{
    {\bf Insights on the physical drivers of ICM turbulence imprinted in the tails of the velocity distributions, according to TNG-Cluster.} We show the mass-weighted distributions of the line-of-sight turbulent velocities in the central regions of TNG-Cluster systems at $z=0$, colour-coded by different cluster properties related to SMBH feedback. So, the $ x $-axis represents the LoS turbulent velocities whereas the $ y $-axis is the fraction of hot gas mass ($ T > 10^{5.5}~\rm{K} $) in each velocity bin. Thin curves denote individual clusters, bold curves show median distributions for each property. Different panels show properties primarily influenced by SMBH feedback (those of as in Fig.~\ref{fig:m-sigma_turb_SMBHProps}): cool-coreness, presence of X-ray cavities, specific star formation rate of the central galaxy, relative (to the average) amount of mechanical energy released by the SMBH. Systems with recent AGN activity exhibit slightly (top) or, in some cases, much broader (bottom) mass distributions of LoS turbulent velocities. That is, clusters with enhanced recent SMBH feedback can contain a small fraction of ICM mass moving at significantly higher turbulent velocities than clusters with more quiescent SMBHs.
  }
  \label{fig:m-Vlos-clusterProp}
\end{figure*}
\subsection{Beyond velocity dispersions}

According to TNG-Cluster, most of the hot gas in cluster centres exhibits subsonic turbulence with $ \mathcal{M} _{\rm{Turb}} < 0.2 $ (90th percentile, Section~\ref{sec:results_machs}). Such low Mach numbers primarily arise from the low turbulent velocities at the cluster centres (Section~\ref{sec:results_distribs}) and not from high sound speeds, as sound speeds in the cluster centres of TNG-Cluster are typically $1000 - 1500 $ km s$^{-1}$. However, small fractions of the ICM mass (of the order per cent) still have supersonic turbulence, and indeed, the distributions of both total and turbulent velocity are very broad (Section~\ref{sec:results_distribs}), even in the central regions.

Based on previous works \citep[chiefly][]{Nelson_2019TNG50, Truong_2024, Ayromlou_2024} and our understanding of the IllustrisTNG SMBH feedback model, we posit that there is a lot of information content in the tails of the velocity distributions, which hence can be harvested to gain insight on the physical processes that shape cluster centres. In \cite{Ayromlou_2024} we showed that the AGN kinetic-mode feedback of IllustrisTNG drives outflows of (and produces) super-virial hot gas at high velocities but such gas has a very small volume and mass fraction, and thus may not necessarily impact the velocity dispersion. Similarly, and also based on TNG-Cluster, in \cite{Truong_2024} we showed that relatively-low inferred velocity dispersions ($100-200$ km s$^{-1}$) do not imply dormant central SMBHs, as feedback can still drive strong bulk flows and high-velocity outflows affecting only a small fraction of the ICM mass (as per our bulk-related findings, Section~\ref{sec:results_distribs}). However, the velocity dispersion does not fully capture the complexity of the ICM velocity field in cluster centres. In this paper, we find that all these subtleties apply also to the turbulent component.

In Fig.~\ref{fig:m-Vlos-clusterProp}, we plot the mass distribution of the LoS turbulent velocities within the central regions of TNG-Cluster systems at $z=0$, colour-coded by different cluster properties related to SMBH feedback -- the same as in Fig.~\ref{fig:m-sigma_turb_SMBHProps}. Thin curves indicate individual clusters; thicker curves show medians for clusters in a given subset.

According to TNG-Cluster, clusters that are strong cool cores, exhibit X-ray cavities, have high star formation rates, or have received more-than-average kinetic-mode SMBH feedback exhibit systematically broader LoS turbulent velocity distributions compared to their counterparts. However this systematic behaviour is more prominent in the tails of the distributions which would not influence the velocity dispersion. In other words, clusters characterized by more recent SMBH feedback activity, have a larger fraction of gas with higher turbulent velocities.

Fig.~\ref{fig:m-Vlos-clusterProp} strengthens our claim that SMBH feedback is a significant driver of turbulence in the ICM of cluster cores (Sections~\ref{sec:results_correlations} and \ref{sec:results_timeevolution}) and suggests that its effects may be more strongly encoded in the tails of the velocity distributions than in their second central moment, i.e. the velocity dispersion. It may therefore be warranted to identify ways to extract information beyond the velocity dispersion from spectral lines, using current or future high-resolution X-ray spectrographs.

\section{Summary and Conclusions}
\label{sec:conclusions}

We have quantified the levels of turbulence in the intra-cluster medium (ICM) of galaxy clusters by analysing the TNG-Cluster suite of cosmological MHD simulations of galaxies (Section~\ref{sec:methods_TNG-Cluster}), thereby accounting for the full cosmological context and a diversity of astrophysical processes, including mechanical feedback from super-massive black holes (SMBHs), as per the IllustrisTNG galaxy-formation model.

We primarily focus on the hot X-ray emitting gas (Section~\ref{sec:methods_defs}) to provide theoretical expectations, at least qualitatively, for the novel observations enabled by high spectral-resolution X-ray spectroscopy, e.g. with \textit{XRISM} and \textit{newAthena}.

{We separate} bulk and turbulent gas motions in the central regions ($\leq 33.5$ kpc, approximately the size of the {\it XRISM} field of view at the redshift of Perseus) of 352 TNG-Cluster systems at $z=0$ spanning $\MTWOC = 10^{14.3-15.4}~\MSUN$, using a multi-scale Reynolds decomposition (RD; Section~\ref{sec:methods_RD}) designed for cell-wise convergence (Fig.~\ref{fig:RD_Schematic}).

Our algorithm successfully separates bulk and turbulent components down to velocities of $10-20$ km s$^{-1}$, de facto resulting in typical turbulence filtering scales of $5-10$~kpc across most ICM gas mass and most clusters (Fig.~\ref{fig:RD_sensitivity}). Visual inspection confirms that the method recovers physically-meaningful structures, such as downstream turbulence in the wakes of SMBH-driven cavities (Fig.~\ref{fig:RD_collection}).\\

Our {main} results, all based on TNG-Cluster predictions, are:

\begin{enumerate}[leftmargin=2em, labelsep=0.75em, labelwidth=1.5em, listparindent=2em]

  \item \textbf{Velocity distributions and Mach numbers:} The velocity distribution of bulk motions is very broad, with high-velocity tails of coherent flows (e.g. inflows and SMBH-driven outflows) that can reach $2000-3000$ km s$^{-1}$ even in the very centres of clusters and with a clear dependence on cluster mass, reflecting the depth of the gravitational potential. In contrast, the distributions of turbulent velocities are narrower, $\mathcal{O}(100)$ km s$^{-1}$, nearly isotropic, and less dependent on cluster mass (Fig.~\ref{fig:vel-distribution}). The turbulent velocity distribution is better represented by a Lorentzian than a Gaussian. Consequently, turbulence in cluster cores is strongly subsonic, with typical Mach numbers $ < 0.2 $ (90th percentile), although small fractions of ICM mass (of order percent for turbulence and ten percent for bulk motions) move supersonically (Fig.~\ref{fig:vel-sound-distribution}).\\

  \item \textbf{Turbulent velocity dispersion:} The turbulent velocity dispersion in cluster centres accounts for about $40-50$ per cent of the total velocity dispersion in the average cluster, and is confined to a narrow range of median values of $50-75$ km s$^{-1}$ across two orders of magnitude in cluster mass, being somewhat higher for cool than for hot X-ray-emitting gas (Fig.~\ref{fig:sigmaturb2tot}).\\

  \item \textbf{Clusters with higher levels of turbulence:} Enhanced turbulent velocity dispersion in cluster centres correlates with recent AGN activity measured as higher-than-average amounts of mechanical energy released by the SMBHs, and indicators of recent feedback such as strong cool-core status, and the presence of X-ray cavities (Fig.~\ref{fig:m-sigma_turb_SMBHProps}).\\

  \item \textbf{Transient turbulence from SMBH feedback:} High time-cadence analysis of the most massive TNG300 cluster reveals that episodes of SMBH kinetic feedback drive, not only outflows up to thousands of km s$^{-1}$, but also short-lived enhancements in both total and turbulent velocity dispersions in the core (by factors of $4-10$, up to $200-500$ km s$^{-1}$). This turbulence dissipates rapidly, returning to pre-injection levels typically within $30-50$ Myr, indicating that SMBH feedback is a key driver of turbulent gas motions but that feedback-induced turbulence is transient (Figs.~\ref{fig:TNG300_subbox0_times},~\ref{fig:TNG300_subbox0_profiles}).\\

  \item \textbf{Radial profiles and pressure support:} The turbulent velocity dispersion, when averaged across all clusters, exhibits a characteristic, `U-shaped' radial profile: peaking in the centre, dipping at intermediate radii ($0.1-0.2\, \RFC$), and rising again towards the outskirts. The non-thermal pressure support derived strictly from turbulence ($P_{\rm Turb}/P_{\rm Therm}$) is below per cent level in the average cluster throughout the cluster volume (Fig.~\ref{fig:radial-prof.sigmas}). \\

  \item \textbf{Cluster-to-cluster variations:}
    Whereas cluster-population averages allow insight into general trends and the dominant underlying physical processes, cluster-to-cluster variations are large. This is the case for the velocity distributions -- especially for the bulk velocity, where individual clusters may or may not exhibit signatures of merger or gas accretion events upon inspection; the values of the turbulent velocity dispersion in cluster centres (with 16th–84th percentiles differing by factors of $2-4$); and the shapes of the 3D radial profiles of turbulent velocity dispersion and turbulent pressure support.\\

\end{enumerate}

From a conceptual perspective, taking together these results support a picture in which SMBH mechanical feedback is an important driver of turbulence in the central regions of galaxy clusters, while concurrently driving high-velocity outflows of order thousands of km s$^{-1}$ for a small fraction of the gas. The dissipation timescales of order tens of million years for feedback-driven turbulence provide a natural explanation for the large cluster-to-cluster variation in the properties of cluster centres, and may also suggest the need for frequent feedback events to maintain turbulent pressure support. In turn, precise measurements of the scatter in velocity dispersion across large cluster samples could help constrain AGN activity and coupling efficiencies.

The stronger dependence on cluster mass of bulk motions compared to turbulence suggests, as expected, that the former are predominantly governed by large-scale gravitational dynamics, with more massive clusters -- having deeper potential wells and growing dominantly via cosmic accretion -- able to drive or host stronger and faster coherent flows. However, also in the turbulent velocities cluster mass dependence manifest, especially in the high-velocity tails. This indicates that turbulence is likely driven by a combination of local processes such as SMBH feedback -- which indirectly depends on the mass of the host galaxy and cluster -- and stronger gas motions in more massive clusters that can also seed turbulence.

Even though our analysis is purely theoretical, without forward modelling into observational space, our findings provide a simulation-based baseline for interpreting current and upcoming high-resolution X-ray spectroscopic observations of galaxy cluster cores. Since turbulence accounts for roughly half of the total velocity dispersion, observational estimates of turbulent energy density based on total line broadening likely overestimate the turbulent energy density by a factor of $\sim 4$, or, in other words, the turbulent pressure support in galaxy clusters is even lower than what is uncovered by current {\it XRISM}  analyses. Spatially-resolved spectroscopy will be essential to separate the underlying components. Furthermore, the large cluster-to-cluster variations predicted by TNG-Cluster highlight the need for caution when drawing strong conclusions from comparisons between individual observed clusters, single simulated systems, or median trends. Finally, our results suggest that the impact of physical processes such as SMBH feedback on ICM turbulence may be more strongly encoded in the high-velocity tails of the velocity distributions than in their second moment. This warrants efforts for observational approaches that go beyond velocity dispersion measurements when interpreting high-resolution X-ray spectroscopic data.

\section*{Acknowledgements}
BS thanks Katrin Lehle, Eric Rohr, and Anirudh Ravishankar for insightful discussions related to the TNG-Cluster simulations and various numerical techniques. The authors also acknowledges Jonathan Stern for insightful discussions on theoretical predictions of cosmological turbulence and Aurora Simionescu on the interpretation of {\it XRISM}  results.

BS is a fellow of the International Max Planck Research School for Astronomy and Cosmic Physics at the University of Heidelberg (IMPRS-HD).
BS, AP, JB, MP and DC acknowledge funding from the European Union (ERC, COSMIC-KEY, 101087822, PI: Pillepich). DN acknowledges funding from the Deutsche Forschungsgemeinschaft (DFG) through an Emmy Noether Research Group (grant number NE 2441/1-1).

The TNG-Cluster simulation suite has been executed on several machines: with compute time awarded under the TNG-Cluster project on the HoreKa supercomputer, funded by the Ministry of Science, Research and the Arts Baden-Württemberg and by the Federal Ministry of Education and Research; the bwForCluster Helix supercomputer, supported by the state of Baden-Württemberg through bwHPC and the German Research Foundation (DFG) through grant INST 35/1597-1 FUGG; the Vera cluster of the Max Planck Institute for Astronomy(MPIA),as well as the Cobra and Raven clusters, all three operated by the Max Planck Computational Data Facility(MPCDF); and the BinAC cluster,supported by the High Performance and Cloud Computing Group at the Zentrum für Datenverarbeitung of the University of Tübingen, the state of Baden-Württemberg through bwHPC and the German Research Foundation (DFG) through grant no INST 37/935-1 FUGG.

All the analysis and computations associated to this paper have been done on the Vera cluster of the MPCDF.

\section*{Data Availability}

The TNG-Cluster simulation data are publicly available at \url{https://www.tng-project.org/cluster/}, following the IllustrisTNG public data release \citep{Nelson_2019}. The data underlying this article, and the analysis code, will be shared upon reasonable request to the corresponding author.



\bibliographystyle{mnras}
\bibliography{refernces}




\appendix
\renewcommand{\thefigure}{\thesection.\arabic{figure}}
\setcounter{figure}{0}  

\section{Velocities and gradients of interest from our Reynolds Decomposition}
\label{sec:methods_scales}

\begin{figure}
  \centering
  \includegraphics[width=0.9\linewidth]{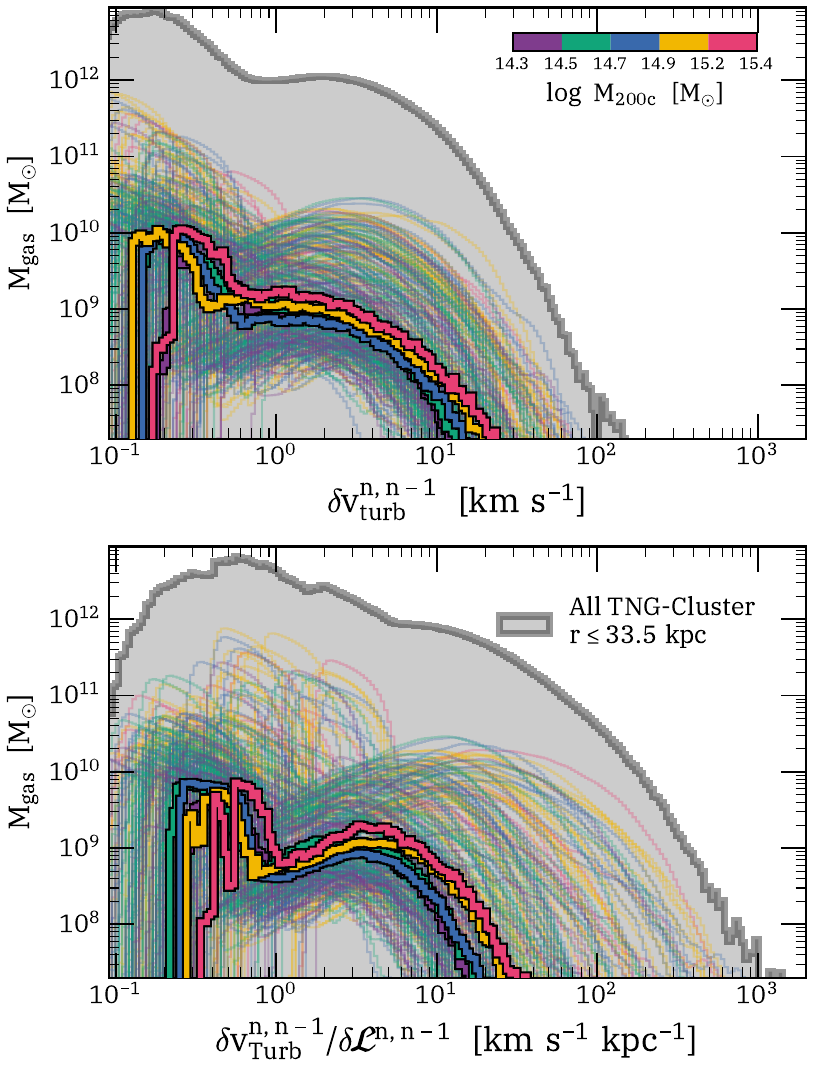}
  \caption{
    {\bf {Sensitivity analysis of the Reynolds Decomposition algorithm for cluster cores in TNG-Cluster.}}
    Panel shows the distribution of two convergence metrics across all gas cells within the central $ 33.5~\rm{kpc} $ sphere of all 352 clusters in TNG-Cluster. Thin curves show individual clusters, bold curves show median profiles in mass bins, and gray shaded regions show total distributions across all simulated clusters.
    \textbf{Top panel:} Distribution of changes in the magnitude of turbulent velocities between final convergence iteration {\tt n} and previous iteration ${\tt n-1}$. This represents the minimum velocity scale that can reliably separable into bulk and turbulent motions ($\sim 10-20$ km s$^{-1}$).
    \textbf{Bottom panel:} Distribution of ratio of change in turbulent velocities to the change in filter length between iteration {\tt [n, n-1]}, indicating the minimum detectable velocity gradients $(\sim 10 ~{\rm km~s^{-1}~kpc^{-1}})$.
  }
  \label{fig:RD_sensitivity}
\end{figure}

The ICM exhibits complex dynamics with multiple coherent motions (e.g., SMBH feedback outflows, {accretion inflows}, merger-induced flows \cite[see e.g.,][]{Simionescu_2019,Ayromlou_2024} that often feature spatial velocity gradients (i.e., $\frac{\mathrm{d}v}{\mathrm{d}r} \neq 0$). While these are technically bulk motions, our algorithm (see Section~\ref{sec:methods_RD}) may not always classify them correctly. This limitation arises from our iterative approach: at each iteration {\tt n}, we compute the local bulk velocity as the average velocity within a spherical region with radius $\mathcal{L}^n$. When velocity gradients are moderate, the changes in our estimates of bulk velocities between iterations may be small enough to satisfy our convergence criteria prematurely, causing some bulk motion to be missed and misclassified as turbulent. Conversely, steep gradients can cause slow convergence. In either case, the recovered bulk and turbulent components may not precisely represent the true physical motions. Furthermore, our spherical averaging inherently smooths the bulk velocity field to some extent. While this effect is less pronounced for uni-directional flows where we can more reliably recover the true bulk motion, the general impact of this smoothing is difficult to quantify precisely.

  {In Fig.~\ref{fig:RD_sensitivity} we attempt to quantify the impact of the unavoidable choices within the multi-scale filtering Reynolds Decomposition adopted throughout. In the top panel, we plot the magnitude of the change in turbulent velocities at the $ [{\tt n, n-1}]^{\rm th} $ iteration i.e $ \delta  v_{\rm{Turb}}^{\rm{n, n-1}} $ , where $ \tt n $ is the iteration where the algorithm converges for a given gas cell. In the bottom panel, we show the analogue distribution for the} change in turbulent velocities with respect to the change in search volume (i.e. search length scale) at the $ [{\tt n, n-1}]^{\rm th} $ iteration. For each panel, thin curves indicate individual clusters; bold curves show mass-bin medians within the {\it XRISM} central pointing, and the gray shaded region shows the total distribution across all the simulated clusters.

  Our algorithm converges whenever the change in {turbulent velocities} for a {given} change in length scale is less than some set threshold (i.e. $ \tau _{\rm{tol}} $); this implicitly sets a minimum velocity and velocity gradient that can be reliably separated into bulk and turbulent motions.

  The top panel can be interpreted as the distribution of minimum velocity that can be reliably separated into bulk and turbulent motions by our algorithm. For our fiducial choice, we find that across all the clusters and all gas in our simulations, the algorithm on average can separate bulk and turbulent motions down to $ \sim 10-20 $ km s$^{-1}$, with {mild} dependence on the cluster mass in the median profiles. We note that while there are gas cells where our measurement are unreliable up-to $ ~ 100 $ km s$^{-1}$ but the fraction of gas mass in those cells is very small (more than 2 orders of magnitude lower than the peak of the distributions).

  The bottom panel can instead be interpreted as the distribution of minimum velocity gradients that can be reliably separated into bulk and turbulent motions by our algorithm. We find that on average we can separate bulk motions with velocity gradients as low as $  \sim 10 $ $\rm km~s^{-1}~kpc^{-1}$.

  \section{Results on ICM turbulence via a fixed filtering length}\label{app:Comparison_FixedFilteringLength}
  \begin{figure*}
    \centering
    \includegraphics[width=1\textwidth]{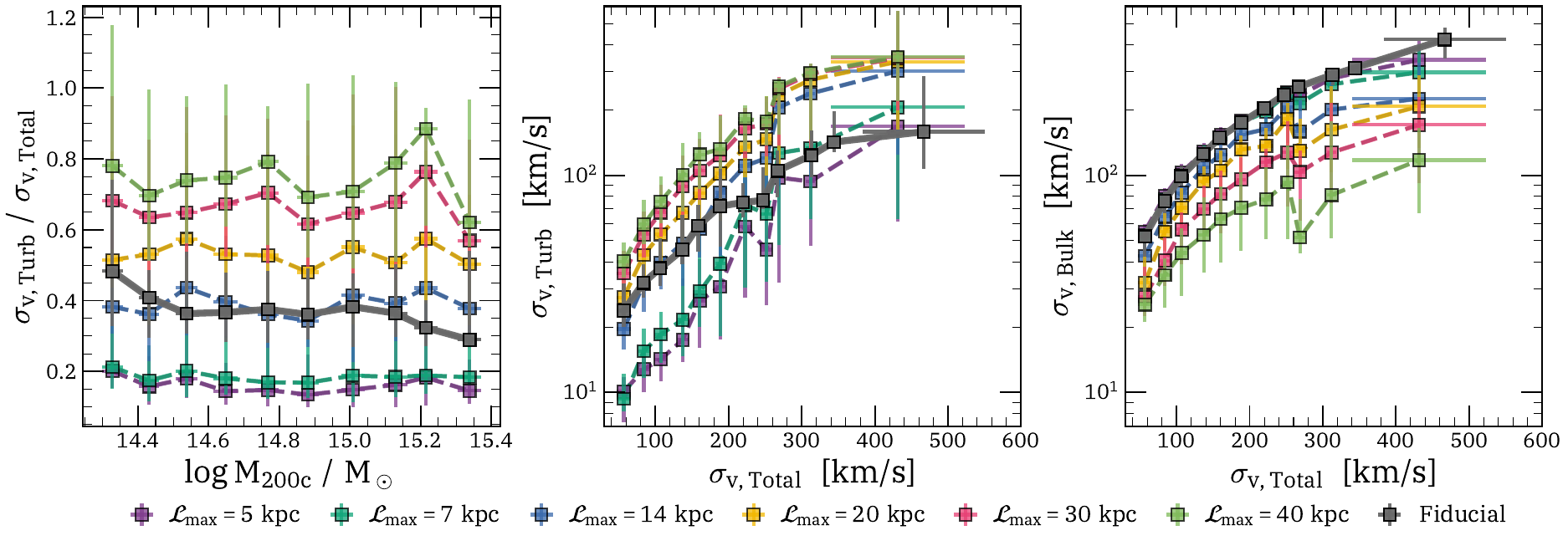}
    \caption{
      {\bf Impact of fixed versus adaptive filtering length on velocity dispersion measurements in TNG-Cluster cores.}
      Each panel shows median profiles (solid lines) with $16^{\rm{th}}$--$84^{\rm{th}}$ percentile error bars for six fixed filtering lengths (5, 7, 14, 20, 30, 40 kpc; coloured lines) compared to the fiducial adaptive multi-scale approach (gray). {\bf Left:} Turbulence fraction ($\svt/\svtot$) versus cluster mass. Larger fixed filtering lengths systematically increase the turbulence fraction by removing more bulk motions. {\bf Middle:} Total (upper curves) versus turbulent (lower curves) velocity dispersion. Fixed filtering lengths progressively overestimate $\svt$ while $\svtot$ remains unchanged. {\bf Right:} Total versus bulk velocity dispersion. Larger filtering lengths underestimate bulk motions. These systematic biases demonstrate that adaptive multi-scale filtering is essential for accurately separating bulk and turbulent velocity components.
    }
    \label{fig:method-FixedFilteringLength}
  \end{figure*}

  Throughout this work, we have used an adaptive multi-scale filtering approach to separate bulk and turbulent velocity fields -- meaning the coherence length scale varies with the local environment. To further highlight the importance of using an adaptive filtering length, in Fig.~\ref{fig:method-FixedFilteringLength}, we show the effects of using a fixed filtering length on the velocity dispersion. We choose six different fixed filtering lengths -- $ 5, 7, 14, 20, 30, 40 $ kpc (shown in different colour) and compare it with the fiducial multi-scale approach (shown in gray). In Fig.~\ref{fig:method-FixedFilteringLength}, the left panel shows the ratio of turbulent to total velocity dispersion, $ \svt / \svtot $, as a function of cluster mass, the middle panel compares the total and turbulent velocity dispersions, while the right panel shows compares the bulk and the total velocity dispersions. In each panel, we only plot the median profiles for the choice of the filtering length, and the error bars represents the $ 16-84^{th} $ percentile range.

  While the simulations do not predict a significant mass trend in the $ \svt / \svtot $ ratio (left panel), we find that  increasing the fixed filtering length leads to a systematic increase in the $ \svt / \svtot $ ratio. This is expected since larger filtering lengths will remove more large-scale bulk motions, thereby increasing the turbulent component. However, this also means that using a fixed large filtering length can lead to an overestimation of the turbulent velocity dispersion. This is further highlighted in the middle panel where we see that increasing the fixed filtering length leads to a systematic increase in the turbulent velocity dispersion, while the total velocity dispersion remains unchanged (since it is independent of the filtering method). In contrast, in the right panel, we see that increasing the fixed filtering length leads to a systematic decrease in the bulk velocity dispersion. Overall, this test highlights the importance of using an adaptive multi-scale filtering approach to separate bulk and turbulent velocity fields -- using a fixed filtering length can lead to biased estimates of both bulk and turbulent velocity dispersions.

  \section{On the functional form of the ICM turbulent velocity distribution and on the meaning of velocity dispersion}
  \label{app:FunctionalForm_TurbVelDist}
  \begin{figure}
    \centering
    \includegraphics[width=\linewidth]{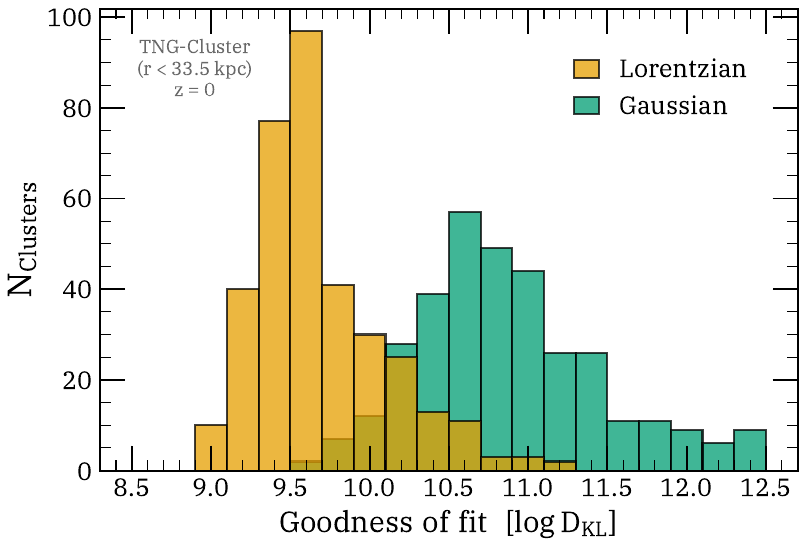}
    \caption{{\bf Summary of the goodness of fit for the mass distribution of LoS turbulent velocities. }
    Yellow and green histograms shows the distribution of Kullback-Leibler (KL) divergence of best-fit Lorentzian and Gaussian profiles respectively, w.r.t  the mass-weighted LoS turbulent velocity distributions of hot gas ($T > 10^{5.5}$ K) within the central regions of TNG-Cluster clusters. Smaller values means better fits. The Lorentzian profile provides a better fit to the data compared to the Gaussian profile, as they better capture the extended high-velocity tails observed in the distributions.}
    \label{fig:gof-hotgas}
  \end{figure}
  The bulk velocity distributions are too diverse across different clusters -- shaped by different physical processes like mergers, feedback and outflows, however, the velocity distribution of the turbulent field in the cluster centres are all very similar across hundreds of clusters, and their symmetric nature suggest that we can fit the distribution with some kind of symmetric profile. Thus we fit these mass distributions of LoS turbulent velocities within the central sphere with Gaussian and a Lorentzian (i.e. Cauchy) profiles. Fig.~\ref{fig:gof-hotgas}, shows the goodness of fit for the Lorentzian and Gaussian fits which is measured by the modified  Kullback-Leibler divergence, $ D_{\rm KL}(Q || P)  $, that quantifies the difference between the observed $ Q(x) $  and true distributions, $ P(x) $ :
  \begin{equation*}
    D_{\rm KL}(Q || P) =
    \begin{cases}
      \sum_x P(x) \log \dfrac{P(x)}{Q(x)} - P(x) + Q(x); \\
      \hspace{25ex} P(x), Q(x)> 0 \\
      Q(x); \hspace{1mm} P(x) = 0, Q(x) > 0  \\
      \infty; \hspace{1mm} \rm{otherwise}
    \end{cases}
  \end{equation*}
  where a smaller value, means better fit \citep{Boyd_Vandenberghe_2004}. For majority of the clusters the Lorentzian profile provides a better fit compared to the Gaussian profiles, indicated by smaller $ D_{\rm KL} $ values. This suggests that the turbulent velocity field in the ICM is better described by a Lorentzian profile. It should be noted that it is the tails of the LoS turbulent velocity distribution that favours a Lorentzian over a Gaussian distribution.

  \begin{figure}
    \centering
    \includegraphics[width=1\linewidth]{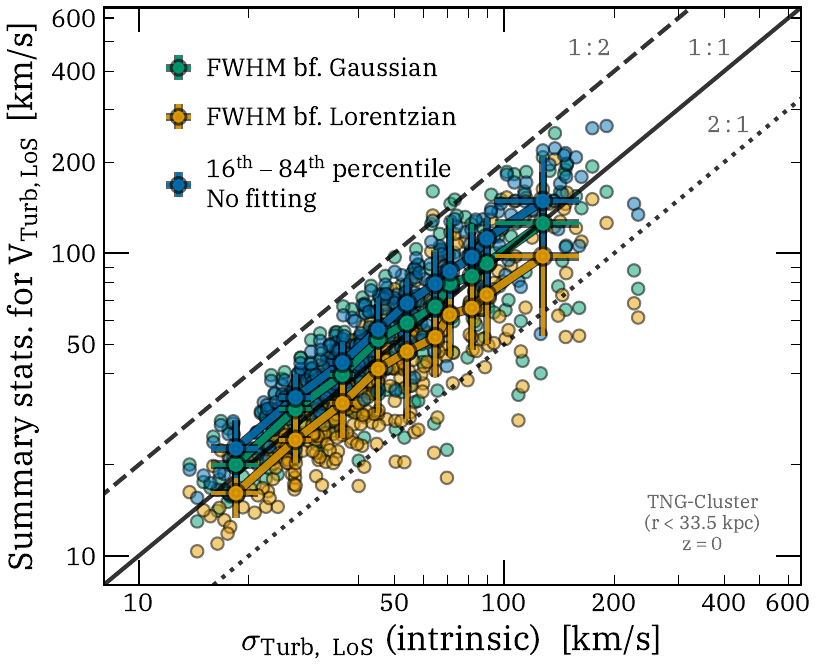}
    \caption{
      \textbf{Gas velocity dispersion of LoS Turbulent velocities $ \left( \sigma_{\rm{Turb, LoS}} \right) $ compared its summary statistics.}
      Blue points show the $16^{\rm{th}}-84^{\rm{th}}$ percentile range from the distribution (without any fitting), yellow and green points show the FWHM of best-fit Lorentzian and Gaussian profiles to the LoS turbulent velocities, respectively. Each point represents one cluster; thick solid lines show median profiles with $16^{\rm{th}}-84^{\rm{th}}$ percentile error bars, and black dashed lines indicate 1:1, 1:2, and 2:1 relations. The FWHM of the best-fit Gaussian most closely traces the intrinsic turbulent velocity dispersion.
    }
    \label{fig:sigma_vs_summaryStats}
  \end{figure}
  Since X-ray observations often assume Gaussian line broadening, we investigate which statistical metric best recovers the intrinsic turbulent velocity dispersion from the line-of-sight distributions. Fig.~\ref{fig:sigma_vs_summaryStats}, shows the intrinsic line-of-sight mass-weighted turbulent velocity dispersion ($x$-axis) against various summary statistics of the line-of-sight turbulent velocity distribution ($y$-axis) within the central regions of TNG-Cluster simulations -- precisely the $ 16^{th}-84^{th} $ percentile range (blue points) and $ 2 \times \gamma $ and $ 2.355 \times \sigma $ which measures the FWHM for the best fit Lorentzian and Gaussian profiles respectively (yellow and green points). Each point represents one cluster in the TNG-Cluster simulations, the thick solid lines represent the median profile with the error-bar representing $ 16^{th}-84^{th} $ range, and the different black lines representing $ 1:1, 1:2 $ and $ 2:1 $ relations.

  Previously established that the Lorentzian profile provides a better fit to the LoS turbulent velocity distributions compared to the Gaussian profile (see Fig.~\ref{fig:gof-hotgas}), but in Fig.~\ref{fig:sigma_vs_summaryStats} we find that the FWHM of the best fit Gaussian better traces the intrinsic LoS turbulent velocity dispersion, $ \sigma _{v, \rm{Turb}} $, compared to the FWHM of the best fit Lorentzian. The median relation between the FWHM of the best fit Gaussian and the intrinsic LoS turbulent velocity dispersion closely follows the $ 1:1 $ line, with a scatter of $ \sim 20-50 $ km s$^{-1}$. On the other hand, the FWHM of the best fit Lorentzian slightly underestimates the intrinsic LoS turbulent velocity dispersion by a factor of $ \sim 0.8 $, with similar scatter of $ \sim 30-60 $ km s$^{-1}$. This suggests that the common assumption of Gaussian velocity distributions remains a valid approximation for the core of the line profile. While the Lorentzian fit better captures the high-velocity tails, the Gaussian fit more accurately recovers the intrinsic velocity dispersion of the bulk of the gas, i.e the core of the distribution.

  We also find that the $ 16^{th}-84^{th} $ percentile range slightly overestimates the intrinsic LoS $ \sigma _{v, \rm{Turb}} $ by a factor of $ ~\sim 1.1 $ for $ \sigma _{v, \rm{Turb, LoS}} < 80$ km s$^{-1}$  with a similar scatter like the best fit Gaussian or Lorentzian points. At the higher values $ \sigma _{v, \rm{Turb, LoS}} $ (i.e. $ > 80 $ km s$^{-1}$) all the three summary statistics tend to show some amount of flattening. This make the $ 16^{th}-84^{th} $ percentile range a more reliable tracer of the intrinsic LoS turbulent velocity dispersion compared  as the median curve now lie on top of the $ 1:1 $ line, whereas the FWHM of the best fit Gaussian or Lorentzian systematically underestimates the $ \sigma _{v, \rm{Turb, LoS}} $.

  \section{Covariance between bulk and turbulent velocity fields}\label{APP:Bulk_Turb_Covar}
  \begin{figure}
    \centering
    \includegraphics[width=1\linewidth]{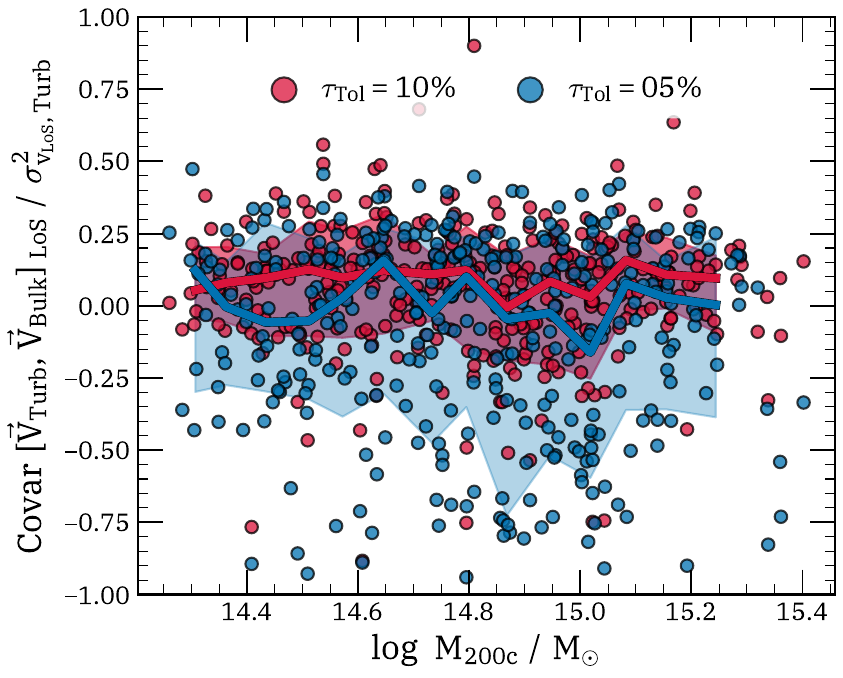}
    \caption{}
    \caption{
      {\bf Negligible covariance between bulk and turbulent velocity components in TNG-Cluster cores.}
      Ratio of the covariance term to the turbulent velocity variance, ${\rm{Covar}}(\vec{v}_{\rm{Bulk}}, \vec{v}_{\rm{Turb}}) / \sigma^2(\vec{v}_{\rm{Turb}})$, for the line-of-sight component as a function of cluster mass. Each point represents one cluster; thick solid lines show median profiles with shaded regions indicating $16^{\rm{th}}$--$84^{\rm{th}}$ percentile ranges. Red shows the fiducial tolerance $\tau_{\rm{Tol}} = 0.1$; blue shows a more stringent value $\tau_{\rm{Tol}} = 0.05$. The median ratio remains near zero across all masses for both tolerances, confirming that the covariance term is negligible. The fiducial choice exhibits tighter scatter ($-0.2$ to $0.2$) compared to the stringent case ($-0.6$ to $0.2$), indicating that overly strict convergence criteria can overestimate bulk velocities. This validates our decomposition $\sigma^2_{\rm{Total}} \approx \sigma^2_{\rm{Bulk}} + \sigma^2_{\rm{Turb}}$.
    }
    \label{fig:Covar.V_LoS}
  \end{figure}

  The key assumption of our method is that for each gas cell, we can write, $ \vec{v}_{\rm{Total}}  = \vec{v}_{\rm{Bulk}} + \vec{v}_{\rm{Turb}} $, which implies
  \begin{align}
    {\sigma^2} \left( \vec{v} _{\rm{Total}} \right) &=  {\sigma^2}\left( \vec{v} _{\rm{Bulk}} \right) +  {\sigma^2}\left( \vec{v} _{\rm{Turb}} \right) \nonumber \\
    &\quad + 2{\rm{Covar}} \left( \vec{v} _{\rm{Bulk}}, \vec{v} _{\rm{Turb}} \right)
  \end{align}
  To validate this assumption, we measure the covariance term directly from the simulations -- for each cluster, we compute LoS covariance using $\vec{v}_{\rm Bulk}$ and $\vec{v}_{\rm Turb}$ as:
  \begin{align}
    \text{Covar}[\vec{v}_{\rm Turb}, \vec{v}_{\rm Bulk}]_{\rm LoS} = &E[\vec{v}_{z, \rm Turb} \cdot \vec{v}_{z, \rm Bulk}] \nonumber\\
    &- E[\vec{v}_{z, \rm Turb}]E[\vec{v}_{z, \rm Bulk}]
  \end{align}
  In Fig.~\ref{fig:Covar.V_LoS}, we plot the ratio of the covariance term to the turbulent velocity dispersion, i.e., $ {\rm{Covar}} (\vec{v} _{\rm{Bulk}}, \vec{v} _{\rm{Turb}}) / {\sigma^2} ( \vec{v} _{\rm{Turb}} ) $ for the LoS component, as a function of cluster mass for the central regions of galaxy clusters in the TNG-Cluster simulation. Each point represents one cluster, the thick solid line represents the median profile, and the shaded region represents the $ 16-84^{th} $ percentile range. Additionally, the different colour represents different choice of tolerance parameter, \( \tau_{\rm{Tol}} \), used in the Reynolds decomposition algorithm -- precisely the fiducial value of \(0.1\) (red), a more stringent value of \(0.05\) (blue). Despite considerable scatter, the median ratios remains very close to zero across the entire mass range for the two different choices of \( \tau_{\rm{Tol}} \). For simplicity, we only show here the LoS component; we have checked for other components and they are similar to the LoS covariance.

  This suggests that on average, covariance term is subdominant compared to the turbulent velocity dispersion. Additionally for our fiducial choice of \( \tau_{\rm{Tol}} = 0.1 \), the scatter is much tighter (lies within $ -0.2, 0.2 $ ) compared to the more stringent choice of \( \tau_{\rm{Tol}} = 0.05 \) (lies within $ -0.6, 0.2 $ ), indicating that going very stringent on the convergence criteria can lead to overestimation of the bulk velocity field, thereby increasing the covariance term. Overall, this test validates our assumption that the covariance term is negligible, justifying our treatment of the total velocity dispersion as the sum of bulk and turbulent components.


  \bsp  
  \label{lastpage}
  \end{document}